\documentclass[a4paper,11pt]{article}
\usepackage{bigfoot}
\usepackage{varioref}
\usepackage{amsmath,amsfonts,amsthm,mathrsfs}

\usepackage[normalem]{ulem}
\usepackage{multirow,color,graphics}
\usepackage{amsmath}
\usepackage{amsfonts}
\usepackage{amssymb}
\usepackage{jheppub}
\usepackage{enumerate}
\usepackage{fancyvrb}
\usepackage{verbatim}
\usepackage{wrapfig}
\usepackage{appendix}
\usepackage{amstext}
\usepackage{amssymb}
\usepackage{graphicx}
\usepackage{color}
\usepackage{varioref}
\usepackage{multirow,graphics}
\usepackage{epstopdf}
\usepackage{simplewick}
\usepackage[makeroom]{cancel}

\usepackage[dvipsnames]{xcolor}
 
\definecolor{Green}{RGB}{0, 142, 0}

\newcommand{\nn}{\nonumber}

\numberwithin{equation}{section}

\def\[{\left[}
\def\]{\right]}
\def\({\left(}
\def\){\right)}

    \newcommand{\beq}{\begin{equation}}
    \newcommand{\eeq}{\end{equation}}

    \newcommand\beqa{\begin{eqnarray}}
    \newcommand\eeqa{\end{eqnarray}}
\newcommand\bea{\begin{array}}
\newcommand\eea{\end{array}}

\newcommand{\bQ}{{\bf Q}}

\newcommand{\bP}{{\bf P}}

\newcommand{\la}[1]{\label{#1}}
\newcommand{\eq}[1]{(\ref{#1})}

\usepackage[dvipsnames]{xcolor}

    \def\bQ{{\bf Q}}
    \def\bP{{\bf P}}

\makeatletter
     \@ifundefined{usebibtex}{} {}
\makeatother

    \def\bQ{{\bf Q}}

        \def\bP{{\bf P}}

\newcommand{\AdS}{\text{AdS}}
\newcommand{\CFT}{\text{CFT}}
\newcommand{\Sphere}{\mathrm{S}}
\newcommand{\Torus}{\mathrm{T}}

\newcommand{\alg}[1]{\mathrm{#1}}

\preprint{DMUS-MP-21/14}

\title{Quantum Spectral Curve for $AdS_3/CFT_2$: a proposal}

\author[a]{~Andrea Cavagli\`a,}
\author[a,b]{~Nikolay Gromov,}
\author[c]{~Bogdan Stefa\'nski, jr.,}
\author[d]{~Alessandro Torrielli}

\affiliation[a]{Mathematics Department, King's College London, The Strand, London WC2R 2LS, UK}
\affiliation[b]{St.Petersburg INP, Gatchina, 188 300, St.Petersburg,
  Russia}
\affiliation[c]{Centre for Mathematical Science, City, University of London, Northampton Square, EC1V 0HB London, UK}  
\affiliation[d]{Department of Mathematics, University of Surrey, Guildford, GU2 7XH, UK}

\emailAdd{andrea.cavaglia${\bullet}$kcl.ac.uk,  nikgromov${\bullet}$gmail.com, Bogdan.Stefanski.1${\bullet}$city.ac.uk,
a.torrielli${\bullet}$surrey.ac.uk}

\abstract{
We conjecture the Quantum Spectral Curve equations for 
string theory on $AdS_3 \times S^3 \times T^4$ with RR charge and its $\CFT_2$ dual. We show that in the large-length regime, under additional mild assumptions, the QSC  reproduces the Asymptotic Bethe Ansatz equations for the massive sector of the theory, including the exact dressing phases found in the literature. The structure of the QSC shares many similarities with the previously known $\AdS_5$ and $\AdS_4$ cases, but contains a critical new feature -- the branch cuts are no longer quadratic. Nevertheless, we show that much of the QSC analysis can be suitably generalised producing a self-consistent system of equations. While further tests are necessary, particularly outside the massive sector, the simplicity and self-consistency of our construction suggests the completeness of the QSC.

}

\begin{document}

\maketitle

\newpage
\setcounter{page}{1}
\section{Introduction}

The Quantum Spectral Curve (QSC) has become an indispensable tool of precision spectroscopy in $\AdS_5/\CFT_4$ and $\AdS_4/\CFT_3$ holographic  models~\cite{Gromov:2013pga,Gromov:2014caa,Gromov:2014bva,Gromov:2015wca,Gromov:2015dfa,Gromov:2015vua,Marboe:2014gma,Marboe:2018ugv,Alfimov:2014bwa,Alfimov:2020obh,Alfimov:2018cms,Gromov:2016rrp,Cavaglia:2014exa,Bombardelli:2017vhk,Gromov:2014eha,Anselmetti:2015mda,Bombardelli:2018bqz,Cavaglia:2018lxi,Grabner:2020nis,Cavaglia:2021bnz}. For a review on the QSC, see~\cite{Gromov:2017blm}. In this paper, we shall take a step towards extending this powerful method to the spectral problem in  another important holographic duality, namely planar $\AdS_3/\CFT_2$. 

It is believed that $\AdS_3/\CFT_2$ dual pairs with 8+8 supersymmetries are  integrable~\cite{Babichenko:2009dk,OhlssonSax:2011ms,Cagnazzo:2012se}.\footnote{For earlier work in this direction see~\cite{David:2008yk,David:2010yg}.} This is the maximal amount of supersymmetry that is allowed for string theory backgrounds of the form $\AdS_3\times \mathcal{M}_7$, with ${\cal M}_7=\Sphere^3\times\Torus^4$ or $\mathcal{M}_7=\Sphere^3\times\Sphere^3\times\Sphere^1$. The symmetries of these two backgrounds are respectively the small and large $(4,4)$ superconformal symmetries, whose Lie sub-algebras are $\alg{psu}(1,1|2)^2$ and $\alg{d}(2,1;\alpha)^2$. The exact S matrices can be found by imposing compatibility with the (centrally extended) vacuum-preserving symmetry algebras of the two theories~\cite{Borsato:2012ud,Borsato:2013qpa,Borsato:2014exa,Borsato:2014hja,Lloyd:2014bsa,Borsato:2015mma}, much like what can be done in higher-dimensional cases~\cite{Beisert:2005tm}. In this paper, we will  focus on string theory on $\AdS_3\times\Sphere^3\times\Torus^4$ supported by R-R charge.

An important difference between these theories and higher-dimensional integrable string backgrounds is the presence of massless excitations in the worldsheet theory, in addition to the more familiar massive ones. The resulting integrable 2-to-2 S matrix breaks up into independent pieces for the scattering of massless/massless, massive/massive and mixed mass excitations. Expressed in terms of Zhukovsky variables, the S matrices  resemble those of higher-dimensional integrable holographic theories, with the mass entering through the shortening conditions. This resemblance is particularly striking in the case of massive excitations~\cite{Babichenko:2009dk,Borsato:2013qpa}, where in the weak-coupling limit the Bethe Equations (BEs) reduce to those of a homogeneous nearest-neighbour $\alg{psu}(1,1|2)\times \alg{psu}(1,1|2)$ spin-chain, with the two factors only connected by the level-matching condition. Away from the weak-coupling limit, the BEs for each $\alg{psu}(1,1|2)$ wing bear a striking similarity to the corresponding part of the $\alg{psu}(2,2|4)$ BEs of $\AdS_5/\CFT_4$. These observations suggest that (at least a part of) the $\AdS_3/\CFT_2$ Q-system can be constructed using two sets of $\alg{psu}(1,1|2)$ Q-functions, one for each wing, and coupling them together in a way that is consistent with the crossing.
The Q-system is an important part of any known QSC \cite{Bombardelli:2017vhk,Gromov:2014caa}. In this paper, instead of deriving the QSC following a long route from TBA equations, we  use the Q-system as a starting point supplying it with the analyticity properties following closely the previously known cases.
However, fairly quickly we realise that one of the analyticity assumptions must be relaxed in our case -- namely we no longer assume the square-root type of singularity near the branch points. This new feature is  inherently connected with the properties of the {\it dressing factors} of~\cite{Borsato:2014hja,Lloyd:2014bsa,Borsato:2015mma}. 

Each  S matrix block comes with a {dressing factor} which is not fixed by symmetry requirements. Dressing factors satisfy crossing equations~\cite{Borsato:2014hja,Lloyd:2014bsa,Borsato:2015mma} that follow from the Hopf algebra structure of the theory~\cite{Janik:2006dc,Gomez:2006va,Plefka:2006ze}. In the case of string theory on $\AdS_3\times\Sphere^3\times\Torus^4$ supported by R-R charge
only, dressing factors which solve these crossing equations have been found~\cite{Borsato:2013hoa,Borsato:2016kbm}. There are two independent dressing phases that enter the massive S matrix, corresponding to either scattering excitations in the same $\alg{psu}(1,1|2)$ wing or in different wings. Their sum is equal to (twice) the Beisert-Eden-Staudacher (BES) phase~\cite{Beisert:2006ez}, while their difference is a new phase, which appears only at the so-called Hernandez-Lopez order~\cite{Beccaria:2012kb}. The relative simplicity of this latter factor is related to the fact that boundstates in the theory can only be made from massive constituent excitations from the same $\alg{psu}(1,1|2)$ wing. As with all solutions of crossing equations, there is potential for CDD ambiguities due to homogeneous solutions of crossing. The absence of such additional factors was demonstrated in~\cite{OhlssonSax:2019nlj}, where it was shown that the proposed dressing factors have exactly the required Dorey-Hofman-Maldacena (DHM) double poles and zeros~\cite{Dorey:2007xn}.

In the case of massive modes, crossing maps the two $\alg{psu}(1,1|2)$ wings into one another.
This suggest that, as a consequence of crossing, the two copies of the 
$\alg{psu}(1,1|2)$ Q-systems should be related by a suitable analytic continuation.
Analogous gluing conditions, which can be traced back to crossing,  
are known to exist in the $\AdS_5/\CFT_4$ and $\AdS_4/\CFT_3$ QSC
and are needed in addition to the QQ-relations to 
constrain the system to a closed system of equations, which can be treated analytically~\cite{Gromov:2015vua,Marboe:2014gma} in some limits and by means of numerical analysis~\cite{Gromov:2015wca} in general. 
Furthermore, the simple gluing of the Q-functions can be shown~\cite{Gromov:2014caa,Bombardelli:2017vhk}
to produce a rather involved expression for the BES dressing phase when considering the large-volume solution.  

While a number of ingredients for the current construction  are borrowed from the known cases,
the new type of near-branch point singularity is a crucial novel ingredient. As a test of our proposal we demonstrated how the ABA equations for the massive sector are precisely reproduced in the large-length limit including the dressing phases. In these considerations, we had to make an additional simplifying assumption about the monodromy of $\mu$-function in the asymptotic limit, which we have not managed to prove. At the same time, we only reproduced the massive sector equations, which suggests that removing this assumption could revive all the massless degrees of freedom, but we leave this question for future work. Another important direction is to verify the completeness of our system of equations by solving it either numerically as in \cite{Gromov:2015wca}  or in a near BPS limits like in~\cite{Gromov:2014bva,Gromov:2014eha}.

An intuitive way in which to understand the effect of massless modes is that the massless dispersion relation can be viewed in an approximate sense as the large coupling limit of the massive one, as long as the particle momentum is kept fixed. 
In the QSC formalism, the coupling usually controls the distance between the cuts in the rescaled spectral  parameter $u/g$. As a result, in the zero mass limit, one might expect this to lead to a number of quadratic cuts collapsing on top of one another. This suggests that, in models with massless modes, the QSC may have a more general singularity structure near the branch points, rather than the conventional square root behaviour seen in higher-dimensional cases. We also expect the analyticity to be simplified in the purely massless sector by employing the pseudo-relativistic variable of~\cite{Fontanella:2019baq,Fontanella:2019ury}.

In fact, the assumption of a square-root singularity is over-restrictive in $\AdS_3$ because it gives rise to a new algebraic constraint on the Q-functions in addition to the QQ-relations. In turn, such a condition collapses the two wings of Q-functions into one, likely leading to drastically simpler analytic properties such as those seen in the Hubbard model~ \cite{Cavaglia:2015nta}, based on a \textit{single} $su(2|2)$ symmetry. 

\vspace{4mm}
The rest of the paper is organised as follows. In section \ref{sec:data}, we collect pre-existing results on integrability for the AdS$_3$/CFT$_2$ duality, which will inspire our conjecture, and describe the algebraic structure of the Q-system for $psu(1,1|2)$. Section \ref{sec:QSC} presents our main proposal for the Quantum Spectral Curve, and describes the unique features of these equations as compared to the previous cases. In section \ref{sec:ABA}, we study the large-volume limit of these equations, reproducing precisely the full Asymptotic Bethe Ansatz for massive modes.  
Finally, we present our conclusions and discuss some future directions. The paper also contains three appendices collecting some notations and  technical details.

\vspace{4mm}
\textit{\bf Note added:} The work described here begun before the epidemic. Shortly after the first wave was coming to an end in Europe, we concluded that the large-length limit was incompatible with square-root cuts as described in section \ref{sec:notquadratic}.  During the recent ``Integrability in Lower Dimensional AdS/CFT" online workshop we learnt that  Simon Ekhammar and Dima Volin had also independently come to a similar conclusion. We are grateful to Dima and Simon for informing us of their findings and coordinating on the release date of the manuscripts to the arXiv. Motivated by these discussions, we revisited our construction and found that relaxing the branch-cut condition allows for a consistent definition of the QSC together with a large-length limit that reproduces the all-loop massive ABA equations found in the literature.
 Our proposal for the QSC seems to be fully consistent with the one  published simultaneously in \cite{Ekhammar:2021pys}.

\vspace{4mm}
\textit{\bf Note added in v3: } 
In the published version of this  article we present a proposal for the QSC, 
whose large-volume solution involves the Riemann-Hilbert problem (\ref{eq:lastRH}). 
In section \ref{sec:dressingrel}, 
the solution to these equations is written in terms of $\chi(u,v)$ functions 
related to the dressing phases proposed in  \cite{Borsato:2013hoa}. These  functions  
 satisfy the correct discontinuity 
equations (\ref{eq:lastRH}), but --  upon closer inspection -- they have an additional branch point at $u\sim \infty$. 
This extra branch point cancels in the full dressing phase of \cite{Borsato:2013hoa}, 
but its presence in the building block $\chi^-(u,v)$ is incompatible with the proposed 
analyticity properties of the QSC. Thus, the claim that our construction reproduces the dressing phases of \cite{Borsato:2013hoa} should be revised. 
\\
After the publication of this paper, a modified proposal for the dressing
phases was made in \cite{Frolov:2021fmj}, for which the extra branch point is absent 
from $\chi^-(u,v)$. We believe that these modified phases have a very good chance of 
arising naturally from our QSC in the large volume limit, at least for the massive-massive 
case,\footnote{We would like to thank Simon Ekhammar, Suvajit Majumder and 
Dmytro Volin for discussions related to this point.}  which would give extra supporting evidence for our construction and at the same time for the conjectured form of the dressing phases. A detailed analysis  will be 
presented elsewhere (for a preliminary discussion see \cite{andreatalk}). \\
The remainder of this arXiv version reproduces in full the published article. 
We emphasize that our proposal for the QSC remains unmodifed, and the only change in the large-volume analysis comes in the explicit form of $\chi^-$.

\section{Data on the AdS${}_3$/CFT${}_2$ integrable system}\label{sec:data}
In this section we assemble together the known facts about the AdS${}_3$/CFT${}_2$ integrable system.
This includes the asymptotic Bethe ansatz for massive modes,
classical algebraic curve and the $psu(1,1|2)$
Q-system.

\subsection{Asymptotic Bethe Ansatz}\label{sec:ABA}
\begin{figure}[t]
    \centering
    \includegraphics[scale=0.4]{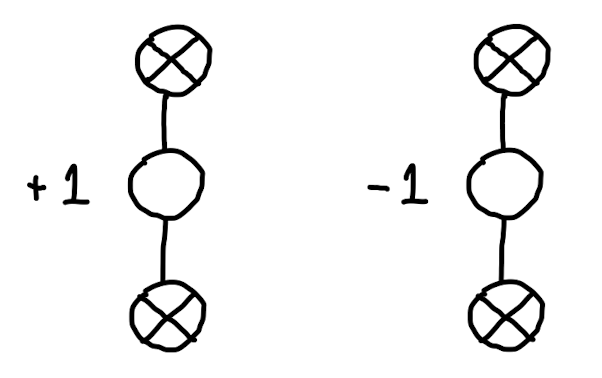}
    \caption{Grading used in the Asymptotic Bethe equations.}
    \label{fig:dyn}
\end{figure}

The massive Asymptotic Bethe Ansatz (ABA) equations which we will be referring to are those presented in~\cite{Borsato:2013qpa}. The symmetry controlling the Bethe equations is $\alg{psu}(1,1|2)^2$. Each copy of $\alg{psu}(1,1|2)$ has associated one momentum carrying root and two auxiliary roots. These are called $x, y_1$ and $y_3$ for one copy of $\alg{psu}(1,1|2)$ and $\bar{x}, y_{\bar{1}}$ and $y_{\bar{3}}$, respectively, for the other copy. The explicit form of the BEs is:
\begin{eqnarray}
1 &=& \prod_{j=1}^{K_2} \frac{y_{1,k} - x_j^+}{y_{1,k} - x_j^-} \prod_{j=1}^{K_{\bar{2}}} \frac{1-\frac{1}{y_{1,k}\bar{x}_j^-}}{1-\frac{1}{y_{1,k}\bar{x}_j^+}},\nonumber\\
\bigg(\frac{x_k^+}{x_k^-}\bigg)^L &=& \prod_{j\neq k}^{K_2} \frac{x_k^+-x_j^-}{x_k^- - x_j^+}\frac{1-\frac{1}{x_k^+ x_j^-}}{1-\frac{1}{x_k^- x_j^+}} \sigma^2(x_k,x_j) \prod_{j=1}^{K_1} \frac{x_k^- - y_{1,j}}{x_k^+ - y_{1,j}} \prod_{j=1}^{K_3} \frac{x_k^- - y_{3,j}}{x_k^+ - y_{3,j}} \nonumber\\
&&\times \prod_{j=1}^{K_{\bar{2}}} \frac{1-\frac{1}{x_k^+ \bar{x}_j^+}}{1-\frac{1}{x_k^- \bar{x}_j^-}} \frac{1-\frac{1}{x_k^+ \bar{x}_j^-}}{1-\frac{1}{x_k^- \bar{x}_j^+}} \tilde{\sigma}^2(x_k,\bar{x}_j) \prod_{j=1}^{K_{\bar{1}}} \frac{1-\frac{1}{x_k^- y_{\bar{1},j}}}{1-\frac{1}{x_k^+ y_{\bar{1},j}}}\prod_{j=1}^{K_{\bar{3}}} \frac{1-\frac{1}{x_k^- y_{\bar{3},j}}}{1-\frac{1}{x_k^+ y_{\bar{3},j}}},\nonumber\\ \la{ABA1}
1 &=& \prod_{j=1}^{K_2} \frac{y_{3,k} - x_j^+}{y_{3,k} - x_j^-} \prod_{j=1}^{K_{\bar{2}}} \frac{1-\frac{1}{y_{3,k}\bar{x}_j^-}}{1-\frac{1}{y_{3,k}\bar{x}_j^+}},
\end{eqnarray}
\begin{eqnarray}
1 &=& \prod_{j=1}^{K_{\bar{2}}} \frac{y_{\bar{1},k} - \bar{x}_j^-}{y_{\bar{1},k} - \bar{x}_j^+} \prod_{j=1}^{K_2} \frac{1-\frac{1}{y_{\bar{1},k}x_j^+}}{1-\frac{1}{y_{\bar{1},k} x_j^-}},\nonumber\\
\bigg(\frac{\bar{x}_k^+}{\bar{x}_k^-}\bigg)^L &=&\prod_{j\neq k}^{K_{\bar{2}}} \frac{\bar{x}_k^- -\bar{x}_j^+}{\bar{x}_k^+ - \bar{x}_j^-}\frac{1-\frac{1}{\bar{x}_k^+ \bar{x}_j^-}}{1-\frac{1}{\bar{x}_k^- \bar{x}_j^+}} \sigma^2(\bar{x}_k,\bar{x}_j) \prod_{j=1}^{K_{\bar{1}}} \frac{\bar{x}_k^+ - y_{\bar{1},j}}{\bar{x}_k^- - y_{\bar{1},j}} \prod_{j=1}^{K_{\bar{3}}} \frac{\bar{x}_k^+ - y_{\bar{3},j}}{\bar{x}_k^- - y_{\bar{3},j}}\nonumber\\
&&\times \prod_{j=1}^{K_2} \frac{1-\frac{1}{\bar{x}_k^- x_j^-}}{1-\frac{1}{\bar{x}_k^+ x_j^+}} \frac{1-\frac{1}{\bar{x}_k^+ x_j^-}}{1-\frac{1}{\bar{x}_k^- x_j^+}} \tilde{\sigma}^2(\bar{x}_k,x_j) \prod_{j=1}^{K_1} \frac{1-\frac{1}{\bar{x}_k^+ y_{1,j}}}{1-\frac{1}{\bar{x}_k^- y_{1,j}}}\prod_{j=1}^{K_3} \frac{1-\frac{1}{\bar{x}_k^+ y_{3,j}}}{1-\frac{1}{\bar{x}_k^- y_{3,j}}},\nonumber\\
1 &=& \prod_{j=1}^{K_{\bar{2}}} \frac{y_{\bar{3},k} - \bar{x}_j^-}{y_{\bar{3},k} - \bar{x}_j^+} \prod_{j=1}^{K_2} \frac{1-\frac{1}{y_{\bar{3},k}x_j^+}}{1-\frac{1}{y_{\bar{3},k} x_j^-}}.
\end{eqnarray}
The Bethe equations are written in the grading illustrated in Fig. \ref{fig:dyn}. 
The massless modes will not be included in our analysis, and they do not feature anywhere in the Bethe equations we write. There is a further level-matching constraint on the solutions to the Bethe equations, in the form of
\begin{eqnarray}
1 = \prod_{j=1}^{K_2} \frac{x_j^+}{x_j^-} \prod_{j=1}^{K_{\bar{2}}} \frac{\bar{x}_j^+}{\bar{x}_j^-}
\end{eqnarray}
(once more disregarding massless modes). The Zhukovsky variables satisfy the familiar constraint given by (suppressing the particle index)
\begin{equation}
x^+ + \frac{1}{x^+} - x^- - \frac{1}{x^-} = \frac{i}{h}, \qquad \frac{x^+}{x^-} = e^{ip},
\end{equation}
where $h$ is the coupling constant of the theory and $p$ is the particle momentum. The same holds for the barred variables. 
The dispersion relation that gives the energy of a particle of momentum $p$ reads
\begin{equation}
E(p) = \sqrt{1 +16 \, h^2 \sin^2 \frac{p}{2}} ,    
\end{equation}
and the anomalous dimension of the state associated to a solution of the ABA is given by 
\beq
\delta\Delta \equiv \gamma =  2 h \sum_{k=1}^{K_2}\left(\frac{i}{x_{k}^+} - \frac{i}{x_{k}^-}\right) + 2 h \sum_{j=1}^{K_{\bar{2}}}\left(\frac{i}{\bar{x}_j^+} - \frac{i}{\bar{x}_{j}^-}\right).
\eeq

The explicit form of the dressing phases from~\cite{Borsato:2013hoa} is given by
\begin{equation}
\sigma(p_1,p_2) = \mathrm{e}^{ i \theta(p_1,p_2)}, \qquad \tilde{\sigma}(p_1,p_2) = \mathrm{e}^{ i \tilde{\theta}(p_1,p_2)},
\end{equation}
with the familiar splitting
\begin{equation}
\theta(p_1,p_2) = \chi(x_1^+,x_2^+)+\chi(x_1^-,x_2^-)-\chi(x_1^+,x_2^-)-\chi(x_1^-,x_2^+),
\end{equation}
with similar expressions for $\tilde{\sigma}$. The individual blocks read
\begin{eqnarray}\la{chidef}
&&
\chi(x,y) = \chi^{\mbox{\scriptsize BES}}(x,y) -\frac{1}{2}\Big[\chi^{\mbox{\scriptsize  HL}}(x,y)-\chi^-(x,y)\Big],
\\ \nn&&
\tilde{\chi}(x,y) = \chi^{\mbox{\scriptsize  BES}}(x,y) -\frac{1}{2}\Big[\chi^{\mbox{\scriptsize  HL}}(x,y)+\chi^-(x,y)\Big].
\end{eqnarray}
The part denoted by BES is the Beisert-Eden-Staudacher~\cite{Beisert:2006ez} dressing phase - its expression can be found reproduced in the review~\cite{Vieira:2010kb}. The same holds for the HL part, referring to the Hernandez-Lopez phase~\cite{Hernandez:2006tk}
\begin{eqnarray}
\chi^{{\rm HL}}(x,y) = \Bigg(\int_{C^+} - \int_{C^-}\Bigg) \frac{dw}{4\pi} \frac{1}{x-w} \Big[ \log (y-w) - \log(y - \frac{1}{w})\Big].
\end{eqnarray}
The new ingredient which was constructed in~\cite{Borsato:2013hoa} is given by
\begin{eqnarray}
\chi^-(x,y) = \Bigg(\int_{C^+} - \int_{C^-}\Bigg) \frac{dw}{8\pi} \frac{1}{x-w} \log \Big[ (y-w)\Big(1-\frac{1}{y w}\Big)\Big] - x \leftrightarrow y,
\end{eqnarray}
where the contours $C^\pm$ denote the upper (resp., lower) half semicircle in the complex $w$-plane, both running anti-clockwise. These expressions are valid in the physical region $|x|>1, |y|>1$. The notation $\chi^-$ is commonly used in the $\AdS_3$ literature for this portion of the phase. The minus sign should not be confused with a shift in the spectral parameter - as will otherwise always be meant in this paper.  

Since we will be merely concerned with the massive modes, it is expected that the Asymptotic Bethe equations which we have written above should be valid exactly in the coupling $h$ but only asymptotically in the length $L$. In other words, wrapping corrections are expected to be exponentially suppressed~\cite{Bajnok:2010ke}. This situation would be rather different were we to include massless modes, whose impact on the TBA is not exponentially suppressed - they are expected to be polynomially suppressed in the presence of mixed massive-massless interactions~\cite{Abbott:2020jaa}, or require exact solutions as in the case of the conformal TBA of~\cite{Bombardelli:2018jkj,Fontanella:2019ury} (see also~\cite{Abbott:2015pps,Dei:2018jyj}). 

Notice also that $4h$
gives the size of the branch cut which goes to zero at weak coupling. Since all interaction between the two $\alg{psu}(1,1|2)$ wings go through the branch-cut, the two wings become completely decoupled in the limit of small coupling constant $h\rightarrow 0$, except for the level-matching condition.  

\subsection{Main features of the classical curve}\label{sec:classical}

The Quantum Spectral Curve is a quantum version of the classical curve, which thus contains crucial structural hints.
We shall from now on denote with un-dotted/dotted indices the variables pertaining to the first/second wing, respectively, of the Dynkin diagram -- corresponding to the first/second copy of $\alg{psu}(1,1|2)$. 

Here we present a short description of some aspects of the algebraic curve describing the integrability of classical solutions of string theory on $AdS_3\times S^3 \times T^4$, following the discussion in~\cite{Babichenko:2014yaa}. 
 This description is based on 4+4 quasimomenta, associated to the fundamental representations of the two $\alg{psu}(1,1|2)$'s. 
They will be denoted by 
$(p_1^A, p_2^A , p_1^{S},p_2^{S})$ and $(p_{\dot 1}^A, p_{\dot 2}^A , p_{\dot 1}^{S},p_{\dot 2}^{S})$. Each quasimomentum naturally parametrises motion in one of the factors of the target space, which is marked by the superscripts $A$, $S$ for $AdS_3$ and   $S^3$, respectively. They are very important quantities which are expected to arise in a WKB-type approximation of the Q-functions in the classical limit of the quantum spectral curve. 

The $p$'s are naturally seen as functions of the Zhukovsky variables, and contain the symmetry charges of the solution in their asymptotics: 
\beqa
\left(\begin{array}{c}
p_1^A\\
p_2^A \\
p_1^{S}\\
p_2^{S}
\end{array}
\right)&\simeq&\frac{1}{2 h x}\left(\begin{array}{c} 
- \Delta-S - \hat{B} \\
+\Delta + S - \hat{B} \\
-J-K-\hat{B}\\
+J+K-\hat{B}
\end{array}
\right)=\frac{1}{2 h x}\left(\begin{array}{c}  -\gamma-2 K_1-L \\
+\gamma+2 K_3 + L \\
-2 K_1+2 K_2-L\\
-2 K_2 + 2 K_3 + L
\end{array}
\right)\;,\\
\left(\begin{array}{c}
p_{\dot 1}^A\\
p_{\dot 2}^A \\
p_{\dot 1}^{S}\\
p_{\dot 2}^{S}
\end{array}
\right)&\simeq&\frac{1}{2 h x}\left(\begin{array}{c}  +\Delta-S - \check{B} \\
- \Delta + S - \check{B} \\
+J-K-\check{B}\\
-J+K-\check{B}
\end{array}
\right)=\frac{1}{2 h x}\left(\begin{array}{c}  +\gamma-2 K_{\dot 1} + 2 K_{\dot 2} + L \\
-\gamma-2 K_{\dot 2}+2 K_{\dot 3} - L \\
-2 K_{\dot 1}+L\\
 2 K_{\dot 3} - L
\end{array}
\right),
\eeqa
where on the rhs we used the explicit expression of the charges  in terms of  Bethe roots numbers. Finally, the classical curve tells us how the quasimomenta in the two wings  are related. In particular, for the quasimomenta describing motion in $AdS_3$, the relation is extremely simple and consists in analytic continuation
\beq\la{clasglue}
p_a^A\left(\frac{1}{x}\right) = p_{\dot a}^A(x), \;\; a=1,2\,,
\eeq
as described in equations (7.13) and (7.38) of~\cite{Babichenko:2009dk}. We will lift this property to the quantum case. 

\subsection{Algebra of the $\alg{psu}(1,1|2)$ Q-system}\label{sec:Qsystem}

The sets of functional relations between the Q-functions (known as Q-systems) take a universal form depending only on the symmetry algebra of the integrable system.  
Since our model contains two copies of $\alg{psu}(1,1|2)$,  important input for our construction comes from the structure of QQ relations for this algebra. 
 
The $\alg{psu}(1,1|2)$ Q-system  contains $16$ independent Q-functions depending on the spectral parameter $u$. They can be labelled  as $Q_{A|I}$, where $A$, $I$ are completely anti-symmetric strings of indices made from $\left\{1,2\right\}$
\beq
A\,, I \in \left\{ \emptyset, 1, 2, (12) \right\}\,,
\eeq
interrelated by the QQ relations 
\beqa\label{eq:QQrel}
 Q_{a A|I} Q_{A|I i}&=&  Q_{aA| I i}^+ Q_{A|I }^- - Q_{aA|Ii}^- Q_{A|I}^+ \,, \label{eq:fermionic}\\
Q_{12|I} Q_{ \emptyset|I} &=& Q_{1|I}^+ Q_{2|I}^- - Q_{1|I}^- Q_{2|I}^+ \,,\\ Q_{A|12} Q_{ A|\emptyset} &=& Q_{A|1}^+ Q_{A|2}^- - Q_{A|1}^- Q_{A|2}^+ \,,\label{eq:bosonic2}
\eeqa
where $a,i \in\left\{1,2\right\}$ are single indices, and $A$, $I$ are anti-symmetric multi-indices defined above. The first type of relation (\ref{eq:fermionic}) is usually called fermionic, and the remaining two bosonic. In these equations, we are using the notation adopted in the whole paper for shifts in the spectral parameter $u$: for any function $g$,
\beq
g^{[\pm n]}(u)\equiv g(u\pm \tfrac{i n}{2})\,, \qquad\qquad g^{\pm}(u)\equiv g^{[\pm 1]}(u).
\eeq
In our proposal, the QSC will contain two copies of these relations, which we will denote by distinguishing between dotted and undotted indices (giving us 16+16 Q-functions). In this section, we focus on one wing, and elaborate on some consequences of (\ref{eq:fermionic})-(\ref{eq:bosonic2}).

We  will make a simple special choice for the Q-functions with the extremal combinations of indices:
\beq\label{eq:unimodular}
Q_{\emptyset|\emptyset} = Q_{12|12} = 1 ,
\eeq
which is analogous to the choice made in the other known QSC cases. Notice that the Q-system has several symmetries, and in particular we are free to set  $Q_{\emptyset|\emptyset} = 1$ through an overall normalisation. The  further, nontrivial algebraic assumption underlying (\ref{eq:unimodular}) is that $Q_{12|12}(u)/Q_{\emptyset|\emptyset}(u)$ is an $i$-periodic function of $u$. Once we have this periodicity property, the analytic properties of Q-functions we will discuss in the next sections imply that  $Q_{12|12}/Q_{\emptyset|\emptyset}$ should be a constant, which we are free to normalise to one using the  symmetries of the Q-system. In quantum spin chains, the periodicity  $1=\frac{Q_{12|12}^+ Q_{\emptyset|\emptyset}^- }{Q_{12|12}^- Q_{\emptyset|\emptyset}^+ }$ can be traced to the quantum transfer matrix having a unit determinant. It is also expected that such condition reflects the projectivity of the algebra $\alg{psu}(1,1|2)$. In particular,  as discussed in \cite{Gromov:2014caa}, it implements a zero-charge constraint for the quantum numbers, which enter the asymptotics of Q-functions in the way described in the next section. For these reasons, from now on we assume the validity of (\ref{eq:unimodular}), which so far seems fully consistent with the description of the $AdS_3$ integrable system. 

We will adopt a  special notation for some of the Q-functions,
\beq
\bQ_k \equiv Q_{\emptyset|k}, \;\;\; \bP_a\equiv Q_{a|\emptyset}, \;\;\; \bQ^k \equiv \epsilon^{kl} Q_{12|l}, \;\;\;\bP^a \equiv \epsilon^{ab} Q_{b|12} ,
\eeq
 as well as $Q^{a|i}\equiv \epsilon^{ab} \epsilon^{ij} Q_{b|j}$. Explicitly,
\beq\label{eq:Qaiupexpl}
Q^{a|i} = \left( \begin{array}{cc} Q_{2|2} & -Q_{2|1}\\
-Q_{1|2} & Q_{1|1} \end{array}\right) ,
\eeq
such that 
\beq
Q_{a|i} Q^{b|i} = \delta_a^b , \;\; \;\;Q_{a|i} Q^{a|j} = \delta_i^j ,
\eeq
due to the unimodularity property
\beq
\text{det}\left( Q_{a|i} \right) = 1 ,
\eeq
which is a consequence of the Q-system with the boundary conditions (\ref{eq:unimodular}). 
Let us write explicitly some of the fermionic equations, which will be used extensively,
\beq\label{eq:fermionic1}
Q_{a|i}^+ -Q_{a|i}^- = \bP_a \bQ_i ,
\eeq
together with $
Q_{a|i}^- -  Q_{a|i}^+ = Q_{12|i} Q_{a|12} 
$, 
which can be rewritten in Hodge-dual notation as
\beq\label{eq:fermionic1dual}
 Q^{a|i \;+} -  Q^{a|i\;-} = -\bP^a \bQ^i .
\eeq
Further useful consequences of the QQ relations are:\footnote{We note that the validity of (\ref{eq:vanishPP}) depends on the constraint $1=\frac{Q_{12|12}^+ Q_{\emptyset|\emptyset}^- }{Q_{12|12}^- Q_{\emptyset|\emptyset}^+}$.}
\beq\label{eq:vanishPP}
\bP_a \bP^a = \bQ_i \bQ^i = 0 ,
\eeq
and the following relations
\beq\label{eq:raisewithQai}
Q_{a|i}^{\pm} \bQ^i = \bP_a \;\;, \;\; Q_{a|i}^{\pm} \bP^a = \bQ_i \;\;,\;\;
Q^{a|i\;\pm}\bQ_i=\bP^a\;\;,\;\;
Q^{a|i\;\pm}\bP_a=\bQ^i\;,
\eeq 
where the equations with $\pm$ signs are compatible due to  (\ref{eq:fermionic1})--(\ref{eq:vanishPP}). 

A useful rewriting of (\ref{eq:fermionic1}), (\ref{eq:fermionic1dual}) incorporating $Q_{a|i}$ is
\beqa\label{eq:keytranslation}
Q_{a|i}^- = Q_{a|j}^+ \left( \delta^j_i -\bQ^j \bQ_i \right), \;\;\; Q^{a|i\;-} = Q^{a|j\;+} \left( \delta_j^i +\bQ^i \bQ_j \right),
\eeqa
or alternatively,
\beqa\label{eq:keytranslation2}
Q_{a|i}^- = Q_{b|i}^+ \left( \delta^b_a -\bP^b \bP_a \right), \;\;\; Q^{a|i\;-} = Q^{b|i\;+} \left( \delta_a^b +\bP^b \bP_a \right).
\eeqa
So far, most of these relations are structurally similar to the ones found for $\alg{psu}(2,2|4)$ - the $AdS_5$ case. 
In this case of lower rank, however, there is an interesting new feature, which follows from the fact that $Q_{a|i}$ and $Q^{a|i}$ are related in a simple manner by~(\ref{eq:Qaiupexpl}). The compatibility of (\ref{eq:fermionic1}) and (\ref{eq:fermionic1dual}) then gives 
\beq
\bQ^k\bP^a=-\epsilon^{kl}\epsilon^{ab}\bQ_l\bP_b ,
\eeq
or explicitly,
\beq
\bQ^1\bP^1=-\bQ_2\bP_2\;\;,\;\;
\bQ^1\bP^2=+\bQ_2\bP_1\;\;,\;\;
\bQ^2\bP^1=+\bQ_1\bP_2\;\;,\;\;
\bQ^2\bP^2=+\bQ_1\bP_1 ,
\eeq
which imply the equalities of  certain ratios of $\bP$
or $\bQ$ functions:
\beq\la{defchi}
\frac{\bQ^1}{\bQ_2}=
-\frac{\bQ^2}{\bQ_1}
=-\frac{\bP_2}{\bP^1}=+\frac{\bP_1}{\bP^2} \equiv r .
\eeq
The quantity $r(u)$ defined above will have an interesting role in our system. Notice that it allows to raise or lower the indices
\beq\la{chi}
\bQ^k=+r \epsilon^{kl}\bQ_l\;\;,\;\;
\bQ_k=-\frac{1}{r} \epsilon_{kl}\bQ^l\;\;,\;\;
\bP^k=-\frac{1}{r} \epsilon^{kl}\bP_l\;\;,\;\;
\bP_k=+r \epsilon_{kl}\bP^l\;.
\eeq
Finally, a useful consequence of the Q-system is the existence of a 2nd order finite difference equation, describing the $\bQ$ functions in terms of the $\bP$ functions (and vice versa). These Baxter-type equations are described in appendix \ref{app:Baxter}. 

\paragraph{Q-system and Bethe ansatz. }
An important consequence of a Q-system is that it immediately implies the existence of Bethe-like equations restricting the positions of the zeros of the Q-functions, which play the role of Bethe roots. In this argument, we anticipate a crucial assumption  on the Q-functions, namely that they do not have any poles. 

One such system of Bethe equations constrains the zeros of the Q-functions
\beq
Q_{\emptyset|1} = \bQ_1, \;\; Q_{1|1}, \;\; Q_{12|1} =- \bQ^2 .
\eeq
For instance, from  (\ref{eq:fermionic1dual}) we learn that
\beq
\left. Q_{1|1}^+ - Q_{1|1}^-\right|_{u \in  \left\{ \text{zeros of } \bQ_1 \right\} } = \left. \bP_1 \bQ_1 \right|_{u \in  \left\{ \text{zeros of } \bQ_1 \right\} } = 0 ,
\eeq
while, since $\bQ_1 \bP_1 = -\bP^2 \bQ^2$, it is also true that
\beq
\left. Q_{1|1}^+ - Q_{1|1}^-\right|_{u \in  \left\{ \text{zeros of } \bP_1 \right\} } = \left. Q_{1|1}^+ - Q_{1|1}^-\right|_{u \in  \left\{ \text{zeros of } \bP^2 \right\} }=  \left. Q_{1|1}^+ - Q_{1|1}^-\right|_{u \in  \left\{ \text{zeros of } \bQ^2 \right\} } = 0.
\eeq
Shifting the bosonic equation $
Q_{1|1}^+ Q_{2|1}^- - Q_{1|1}^- Q_{2|1}^+ = -\bQ^2 \bQ_1
$ by $\pm i/2$, we also obtain
\beqa
\left. Q_{1|1}^{++} Q_{2|1} \right|_{u \in  \left\{ \text{zeros of } Q_{1|1} \right\} }  &=& \left. - \bQ^{2 +}\bQ_1^+   \right|_{u \in  \left\{ \text{zeros of } Q_{1|1} \right\} } ,\label{eq:toratio1} \\
\left. Q_{1|1}^{--} Q_{2|1} \right|_{u \in  \left\{ \text{zeros of } Q_{1|1} \right\} }  &=& \left. + \bQ^{2 -}\bQ_1^-   \right|_{u \in  \left\{ \text{zeros of } Q_{1|1} \right\} } .\label{eq:toratio2}
\eeqa
The above constraints can be recast as the exact Bethe  equations\footnote{In the case where the Q-functions have cuts, such as will be our system, the relation will be valid on the main Riemann sheet where the Q-system is defined.}
\beqa\label{eq:BAasyL}
&&\left. \frac{Q_{1|1}^{+} }{Q_{1|1}^{-} }  \right|_{u \in  \left\{ \text{zeros of } \bQ_1  \right\} } = 1 \label{eq:ABA1L}\\
&&\left. \frac{Q_{1|1}^{++ } \bQ_1^{-} \bQ^{2\,-} }{Q_{1|1}^{-- } \bQ_1^{+} \bQ^{2\,+}  } \right|_{u \in \left\{\text{zeros of } Q_{1|1} \right\} } = - 1 , \label{eq:ABA2L}\\
&&\left. \frac{Q_{1|1}^{+}   }{Q_{1|1}^{-}  } \right|_{u \in \left\{\text{zeros of } \bQ^2 \right\} } = 1 , \label{eq:ABA3L}
\eeqa
where the middle relation comes from the ratio of  (\ref{eq:toratio1}),(\ref{eq:toratio2}). In a similar way one can deduce several other systems of Bethe equations. For instance, relations of the same form are valid for the zeros of the functions $\bP_1$, $Q_{1|1}$, $\bP^2$. We write these relations with a dot, anticipating that they will be relevant for the second wing:
\beqa\label{eq:BAasyR}
&&\left. \frac{Q_{\dot 1|\dot 1}^{+}  }{Q_{\dot 1| \dot 1}^{-} }  \right|_{u \in  \left\{ \text{zeros of } \bP_{\dot 1}  \right\} }  = 1 \label{eq:ABA1Lb} \\
&& \left. \frac{Q_{\dot 1|\dot 1}^{++} \bP_{\dot 1}^{-} \bP^{\dot 2 \, -} }{Q_{\dot 1|\dot 1}^{--} \bP_{\dot 1}^{+} \bP^{\dot 2 \, +} }  \right|_{u \in  \left\{ \text{zeros of } Q_{\dot 1|\dot 1}  \right\} }  = - 1 , \label{eq:ABA2Lb}\\
&&\left. \frac{Q_{\dot 1|\dot 1}^{+} }{Q_{\dot 1|\dot 1}^{-}  }  \right|_{u \in  \left\{ \text{zeros of } \bP^{\dot 2}  \right\} }  = 1 , \label{eq:ABA3Lb}
\eeqa
In a system like the ones arising in AdS/CFT, the Q-functions are in general complicated functions not known explicitly, therefore such exact Bethe equations have limited practical usefulness when analysing generic solutions of the QSC. However, for certain classes of solutions, such as those with large charges or near special points in the moduli space of the holographic theory the Q-functions do simplify. In the last section of the paper, we find  the explicit large-volume limit of some Q-functions, arising from our QSC equations. Exact Bethe equations such as the ones given above will then reduce to the ABA equations. Additionally, AdS${}_3$/CFT${}_2$ dual pairs have multiple moduli, which preserve integrability~\cite{OhlssonSax:2018hgc} and at special points in the moduli space of each holographic pair additional simplifications to the exact Bethe equations may occur. For example, the weakly coupled RR-charged theory is expected to describe a nearest-neighbour integrable spin chain~\cite{OhlssonSax:2014jtq}.

\section{Proposal for the  QSC}\label{sec:QSC}
In this section we describe the structure of the proposed Quantum Spectral Curve for $\AdS_3$. In the absence of the general TBA equations we cannot follow the usual route of \cite{Cavaglia:2010nm,Gromov:2011cx,Gromov:2014caa} to derive the QSC from TBA. Instead we will be guided by the common properties of the known QSCs for $\AdS_5$
and $\AdS_4$. 

If we summarise the known QSCs there are two main ingredients: QQ-relations, and analytical properties of Q-functions. We consider these components in turn in the following. 

\subsection{Introducing the Q-functions}
\paragraph{QQ-relations. }
In the known case, the QQ-relations follow from the structure of the symmetry of the system. In $\AdS_3$ we have two copies of $\alg{psu}(1,1|2)$
and a natural assumption would be to have two copies of QQ-relations for $\alg{psu}(1,1|2)$,
described in the previous section. To distinguish the two copies we will use dotted indices for one of them, so we will use the following sets of indices ($a=1,2,\;k=1,2$ and same for dotted indices)
\beqa
\bQ_k,\;\bP_a,\;Q_{a|k}
&\leftrightarrow& Q^{a|k},\;\bQ^k,\;\bP^a\;,\\
\bQ_{\dot k},\;\bP_{\dot a},\;Q_{\dot a|\dot k}
&\leftrightarrow& Q^{\dot a|\dot k},\;\bQ^{\dot k},\;\bP^{\dot a}\;.
\eeqa
The above Q-functions are related by the QQ-relations. A
distinguished subset of them, from which one can recover the remaining Q-functions are
\beq
\bP_a\;,\;\bP^a\;\;{\rm and}\;\;
\bP_{\dot a}\;,\;\bP^{\dot a}\;\;{\rm constrained\; by}\;\;\bP_a \bP^a=\bP_{\dot a}\bP^{\dot a}=0\;.
\eeq
For example, $\bQ_k$ can be reconstructed from $\bP_a$ and $\bP^a$ by solving the second order finite-difference equation 
\beq\la{Bax}
\bQ_k^{++}
D_1^-
-
\bQ_k
D_2
+
\bQ^{--}_k
D_1^+=0\;,
\eeq
with the coefficients depending solely on $\bP$'s:
\beq\la{Ds}
D_1=
\epsilon_{ab}
\bP^{a-}\bP^{b+}\;\;,\;\;
D_2=
\epsilon_{ab}
\bP^{a--}\bP^{b++}
-
\bP_c \bP^{c--}\;
\epsilon_{ab}
\bP^{a}\bP^{b++}\;.
\eeq
The above relation, derived in appendix \ref{app:Baxter}, is a consequence of the QQ-relations, so
an identical equation holds for the dotted Q-functions.
Equally one can interchange $\bQ\leftrightarrow\bP$ in \eq{Bax} and \eq{Ds}.

\paragraph{Classical correspondence. }

In the classical limit, described by strong coupling $h\rightarrow \infty$ and large quantum numbers scaling as  $\sim h$, we expect that the quasimomenta appear in a WKB approximation of some of the Q-functions.  In particular, they should be directly related to the Q-functions living in the fundamental representation of each $\alg{psu}(1,1|2)$ algebra. With the notation borrowed from the other cases, we link $\bP$'s with the quasi-momenta associated with $\Sphere^3$  and $\bQ$'s with the ones for $\AdS_3$. 

For the first wing, we will take this correspondence to be the following:
\beqa\label{eq:classical1}
&&\left( \bQ_1 , \bQ_2 | \bP_1 ,\bP_2 \right) \sim  \left( e^{- \int^u {p}^A_1 } , e^{- \int^u {p}^A_2 } | e^{ + \int^u {p}^S_1 } , e^{+ \int^u {p}^S_2 } \right) 
,\\
&& \left( \bQ^1 , \bQ^2 | \bP^1 , \bP^2 \right) \sim  \left( e^{ \int^u {p}^A_1 } , e^{ \int^u {p}^A_2 } | e^{ - \int^u {p}^{S}_1 } , e^{- \int^u {p}^S_2 } \right) , 
\eeqa
which is structurally the same as in $\AdS_5$. 
For the second wing, we take\footnote{Comparing (\ref{eq:classical1}) and (\ref{eq:classical2}), the reader will notice that we reordered some of the labels in the second wing. This is just an arbitrary choice with no loss of generality at this stage (notice that $1\leftrightarrow 2$ in the indices is a trivial symmetry of the Q-system), but it will be convenient for the future, as it  will make the discussion more symmetric between the two wings. }
\beqa\label{eq:classical2}
&&\left( \bQ_{\dot 1} , \bQ_{\dot 2} | \bP_{\dot 1} , \bP_{\dot 2} \right) \sim  \left( e^{ - \int^u  {p}^A_2 } , e^{- \int^u {p}^A_1 } | e^{ \int^u  {p}^S_2 } , e^{\int^u {p}^S_1 } \right),\\
&&\left( \bQ^{\dot 1} , \bQ^{\dot 2} | \bP^{\dot 1} , \bP^{\dot 2} \right) \sim  \left( e^{  \int^u p^A_{\dot 2} } , e^{ \int^u {p}^A_{\dot 1} } | e^{ -\int^u {p}^S_{\dot 2} } , e^{-\int^u {p}^S_{\dot 1} } \right). \label{eq:classical4}
\eeqa
\paragraph{Large-$u$ asymptotics. }
Consistently with the quasi-classical identifications~\eq{eq:classical4} and the asymptotics of the quasimomenta described in section \ref{sec:classical}, the Q-functions should exhibit  power-law asymptotics at large $u$, with behaviour characterised by the charges. In particular, we assume
\beq
\label{eq:largeu}
\bP_a\simeq A_a u^{M_a} , \;\;\; \bP^a \simeq A^a \, u^{-M_a - 1}, \;\;\;\bQ_i\simeq B_i u^{\hat{M}_i} , \;\;\; \bQ^i\simeq B^i u^{-\hat{M}_i-1} ,  
\eeq
for large $u$, where
\begin{align}
M_a&\equiv \left(- \frac{L}{2}  + K_2 - K_1 -1 , \;\frac{L}{2} - K_2 + K_3 \right), & \hat{M}_k &\equiv \left(\frac{\gamma}{2}  + \frac{L}{2}  + K_1 ,\; -\frac{\gamma}{2}   - \frac{L}{2} - K_3 -1 \right),
\end{align}
\begin{align}
M_{\dot a}&\equiv \left(- \frac{L}{2}  +  K_{\dot 3}, \;\frac{L}{2} - K_{\dot 1} -1  \right), & \hat{M}_{\dot k} &\equiv \left(\frac{\gamma}{2}  + \frac{L}{2}  + K_{\dot 2} - K_{\dot 3} -1,\; -\frac{\gamma}{2}  - \frac{L}{2} - K_{\dot 2} + K_{\dot 1} \right).
\end{align}
In the following sections, we will see that some of the Q-functions have horizontal cuts connecting to infinity. In this case, the asymptotic behaviour above will be assumed to be valid for $\text{Im}(u)>0$. 

Notice that the classical identification  is valid in a regime of large quantum numbers, so that it only fixes the structure of (\ref{eq:largeu}) up to finite shifts. However, those can be fine-tuned by the match with the ABA which will be described in the last section of the paper. We will take the exact asymptotics of the Q-functions to be as above. 

\paragraph{Constraints on the constant prefactors and shortening conditions. }
The pre-factors $A$ and $B$ in $\bP$ and $\bQ$ functions
\eq{eq:largeu} usually play an important role.
They can be determined by plugging the large $u$ expansion into the QQ-relations or Baxter equation.
This leads to the following identities
\beq
 r_0\;h^{\hat B}\frac{\prod(-y_{\dot 3,i})}{\prod(-y_{\dot 1,i})}=\frac{B^1}{B_2}=\frac{A_1}{A^2}\,\qquad \quad
 r_0\;h^{\check B}\frac{\prod(-y_{ 3,i})}{\prod(-y_{ 1,i})}=\frac{B^{\dot 1}}{B_{\dot2}}=\frac{A_{\dot1}}{A^{\dot2}}\;.
\eeq
The Baxter equation then implies
\beqa
B_1 B^1=-B_2 B^2=\frac{i}{4}\frac{(\Delta-J-K+S)(\Delta+J+K+S+2)}{\Delta+S+1},\\
A_1 A^1=-A_2 A^2=\frac{i}{4}\frac{(\Delta-J-K+S)(\Delta+J+K+S+2)}{J+K+1},
\eeqa
and with dots
\beqa
B_{\dot 1} B^{\dot 1}=-B_{\dot 2} B^{\dot 2}=\frac{i}{4}\frac{(\Delta-J+K-S)(\Delta+J-K-S-2)}{\Delta-S-1},\\
A_{\dot 1} A^{\dot 1}=-A_{\dot 2} A^{\dot 2}=\frac{i}{4}\frac{(\Delta-J+K-S)(\Delta+J-K-S-2)}{J-K-1}.
\eeqa
 Above we used the following relation between the charges and the Bethe root numbers:
\begin{equation}
    \begin{aligned}
 \Delta &=\gamma+L+K_{\dot{2}}
   +\frac{K_1}{2}+\frac{K_3}{2}-\frac{K_{\dot{1}}}{2}-\frac{K_{\dot{3}}}{2
   }\;, \\
 S&=\frac{K_1}{2}+\frac{K_3}{2}+\frac{K_{\dot{1}}}{2}+\frac{K_{\dot{3}}}{2}-K_{\dot{2}}\;,
   \\
 K&=\frac{K_1}{2}+\frac{K_3}{2}+\frac{K_{\dot{1}}}{2}+\frac{K_{\dot{3}}}{2}-K_2\;, \\
 J&=L-K_2+\frac{K_1}{2}+\frac{K_3}{2}-\frac{K_{\dot{1}}}{2}-\frac{K_{\dot{3}}}{2}\;, \\
  \hat{B}&=K_1-K_3\;, \\
 \check{B}&=K_{\dot{1}}-K_{\dot{3}}\;.
\end{aligned}
\end{equation}
The half-BPS shortening condition $\Delta=J$ and $S=K$ follows from requiring for $A$ and $B$ to vanish. This is an integrability-based derivation of a non-renormalization result for theories with small $(4,4)$ super-conformal symmetry. In such theories, there are left or right sub-algebra shortening conditions: $\Delta_L=J_L$ or $\Delta_R=J_R$. It is well-known that at generic points in the moduli space states which are short with respect to only one such sub-algebra (\textit{i.e.} quarter-BPS states) are not protected, while states which satisfy both shortening conditions (half-BPS states) do not receive quantum corrections~\cite{deBoer:1998kjm,deBoer:1998us,Baggio:2012rr}. An independent derivation of these results was found using ABA methods~\cite{Baggio:2017kza,Majumder:2021zkr} which are valid in the large $L$ limit. The QSC derivation presented here, showing that only half-BPS states are protected, is valid for all lengths $L$.

\subsection{Analytic properties}
\begin{figure}[t]
    \centering
    \includegraphics[scale=0.4]{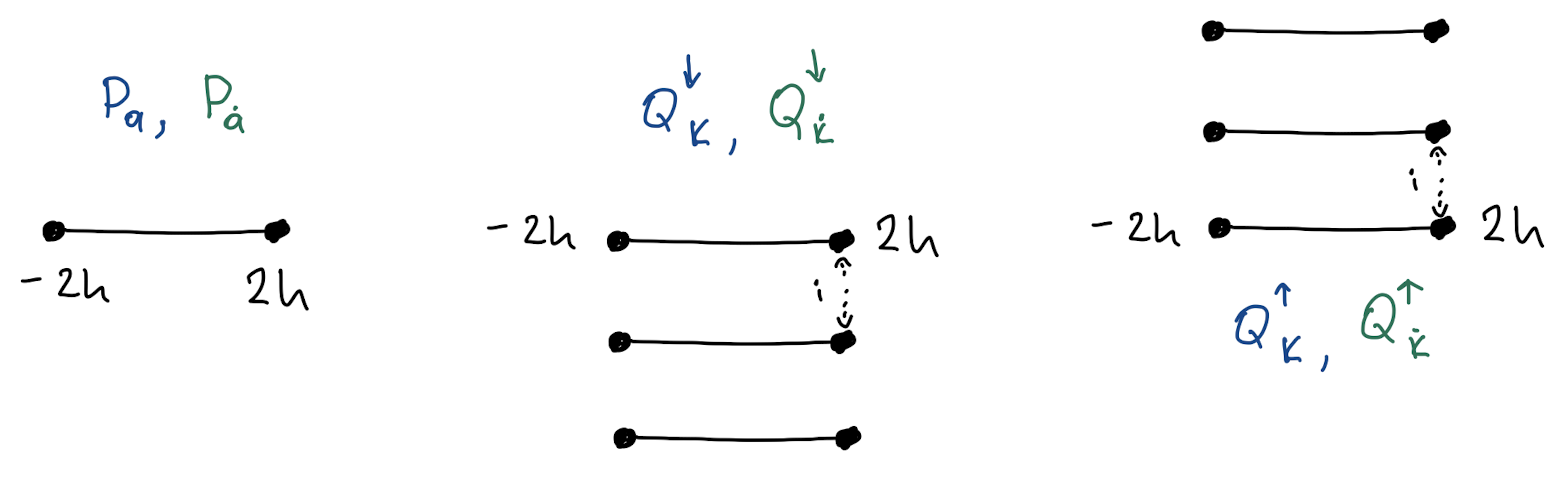}
    \caption{Standard analytic structure of $\bP$'s with one branch cut. As a consequence of this, $\bQ$ functions will have an infinite ladder of cuts separated by $i$ in the lower or upper half of the analytic plane.}
    \label{fig:pq}
\end{figure}
As in all other studied cases, we assume that all $4$ types of $\bP$'s have only one branch cut $(-2h,2h)$ on the real axis and no other singularities on either sheet of their Riemann surface, as shown on Figure~\ref{fig:pq}. Since the $\bQ$-functions are determined in terms of $\bP$'s by means of equation~\eq{Bax}, the analytic properties of $\bQ$ can be deduced from those of $\bP$.
Before describing them let us introduce two different bases of solutions of $\eq{Bax}$: 
\beqa
\bQ_k^\downarrow\;\;&-&\;\;{\rm Upper\;Half\;Plane\;analytic\;(UHPA)\;solutions}\\
\bQ_k^\uparrow\;\;&-&\;\;{\rm Lower\;Half\;Plane\;analytic\;(LHPA)\;solutions}\;.
\eeqa
As the coefficients of \eq{Bax} only have a few cuts near the real axis, and are analytic otherwise, we can always find two solutions of \eq{Bax} $\bQ_k^\downarrow$ which do not have cuts in the UHP, and another pair of solutions $\bQ_k^\uparrow$ which are analytic in the LHP.
Rewriting \eq{Bax} as
\beq\la{Bax2}
\bQ_k^\downarrow
=\frac{\bQ^{\downarrow++}_k
{\color{orange}D^{++}_2}
-
\bQ_k^{\downarrow[+4]}
{\color{orange} D_1^+}
}{D_1^{[+3]}}
\;,
\eeq
and assuming that $\bQ_k^\downarrow$ is analytic for ${\rm Im}\;u>0$ we see that the highlighted terms in the r.h.s. will produce a branch cut on the real axis. Iterating further \eq{Bax2} with shifts $u\to u-2 i n$ in general we generate a ladder of cuts going down the complex plane like on Figure~\ref{fig:pq}.

At the same time, since there are only two linearly independent (with periodic coefficients) solutions of a second order equation \eq{Bax} there must exist an $i$-periodic function (with short cuts) $\Omega_{k}^{\;\;l}$ which relates the two sets of solutions
\beq\label{eq:defOmega}
{\bQ}^\uparrow_k = \Omega_k^{\;\;m} \bQ^\downarrow_{m}\;\;,\;\;
\Omega_k^{\;\;m}(u+i)
=
\Omega_k^{\;\;m}(u)\;.
\eeq
In fact one can write $\Omega_k^{\; m}$ explicitly in terms of $\bQ$'s
\beq
\Omega_k^{\;\; m}=\epsilon^{ml}\frac{
\bQ^{\uparrow}_k
\bQ^{\downarrow--}_l-
\bQ^{\uparrow--}_k
\bQ^{\downarrow}_l
}{
\bQ^{\downarrow}_1
\bQ^{\downarrow--}_2-
\bQ^{\downarrow--}_1
\bQ^{\downarrow}_2
}\;
\eeq
and the periodicity can be verified using \eq{Bax}. There are identical equations for the dotted indices. Furthermore, 
in Appendix \ref{app:Baxter} we show that 
the Hodge-dual Q-functions also satisfy
\beq\label{eq:upOmega}
\bQ^{\uparrow k} = \Omega^k_{\;\;m} \bQ^{\downarrow m}\,\qquad
\Omega^k_{\;\;m}\Omega_k^{\;\;l}=\delta^l_m\;.
\eeq

\vspace{.5cm}

\paragraph{Gluing conditions.}
So far, the two Q-systems were existing independently.
Here we propose a particular way of joining them together.
\begin{figure}
    \centering
    \includegraphics[scale=0.5]{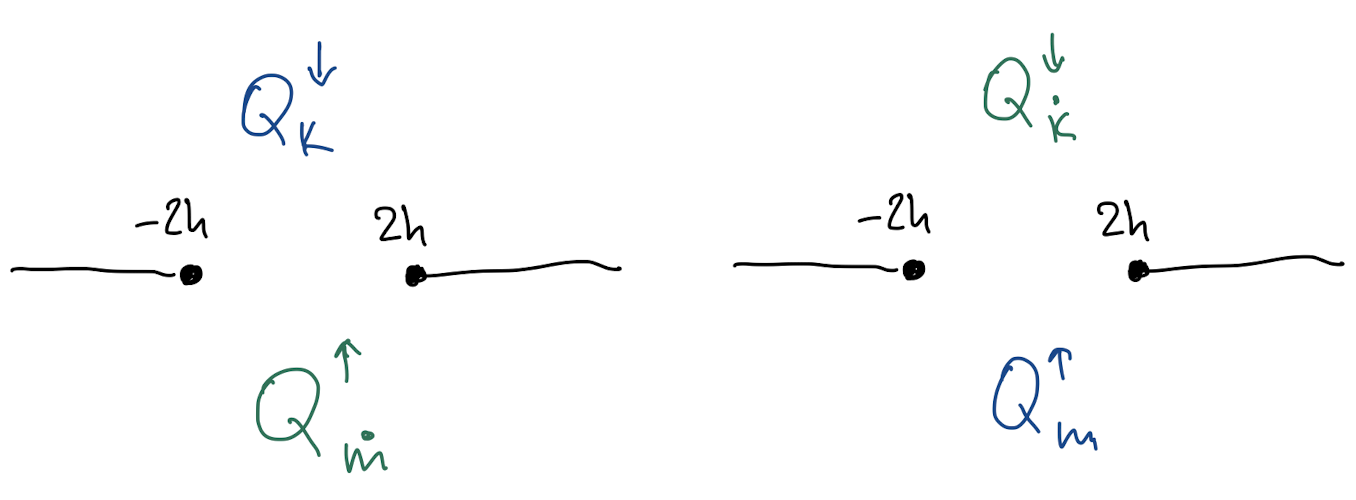}
    \caption{Two $\bQ$'s from different Q-systems are glued together.}
    \label{fig:my_label}
\end{figure}
The underlying idea is to fix the apparent
asymmetry between the analytic properties
of $\bQ$ and $\bP$ (see Figure~\ref{fig:pq}).
Whereas $\bP$ has only one branch-cut, as we argued above, $\bQ$ should have a ladder of cuts going either up or down from the real axis. Following the observation in other QSCs, we notice that a section of the Riemann surface of $\bQ$'s with long cut i.e. $(-\infty,-2h)\cup(2h,\infty)$ on the real axis should not have any other cuts. More specifically we require that (see Fig. \ref{fig:my_label})
\beq\la{gluing1}
\bQ_k^\downarrow(u+i0)=G_k^{\;\;\dot n}\;\bQ_{\dot n}^\uparrow(u-i0)\;\;,\;\;
\bQ_{\dot k}^\downarrow(u+i0)=G_{\dot k}^{\;\; n}\;\bQ_{n}^\uparrow(u-i0)\;\;,\;\;u\in (-2h,2h)
\eeq
where $G_k^{\;\;\dot n}$ and $G_{\dot k}^{\;\;n}$ are two different independent constant matrices. In the studied cases of QSC they have several zero components, but in our case their exact form  is still to be deduced. However, one can make a first guess by looking at the classical counterpart of the gluing relations \eq{clasglue}. Using the identification \eq{eq:classical1} we see that it suggests $G_{1}^{\;\dot 2}$
and $G_{2}^{\;\dot 1}$ to be the only non-zero elements of $G_{k}^{\;\;\dot n}$.

For the Hodge-dual Q-functions, the gluing conditions take a similar form
\beq\la{gluing2}
\bQ^{k\downarrow}(u+i0)=G^k_{\;\;\dot n}\;\bQ^{\dot n\uparrow}(u-i0)\;\;,\;\;
\bQ^{\dot k\downarrow}(u+i0)=G^{\dot k}_{\;\; n}\;\bQ^{n\uparrow}(u-i0)\;\;,\;\;u\in (-2h,2h).
\eeq
Like in the known cases, we assert that gluing is a symmetry of the Q-system 
\beq
G^k_{\;\;\dot n} = \epsilon^{kl}\epsilon_{\dot n\dot m} G^{ \;\;\dot m}_{l}\;\;,\;\;
G^{\dot k}_{\;\;n} = \epsilon^{\dot k\dot l}\epsilon_{n m} G^{\;\; m}_{\dot l}\;\;,\;\;\det G=1\;.
\eeq
In the following, we will choose a basis of Q-functions with specified large-$u$ asymptotics on the first sheet, described in \eq{eq:largeu}. after this choice is made, we are not free to diagonalise the gluing matrix with a  linear transformation. For this reason, we will keep track of it explicitly throughout.   We leave for future work the discussion of the matrix structure of $G$ in this special basis, but as we argued above the classical limit suggests an off-diagonal structure for this matrix.

\vspace{0.5cm}

\paragraph{Properties of the $r$-function.}
The $r$-function, which was defined in section \ref{sec:Qsystem} and allows to lower and raise indices, has interesting analyticity properties. From~\eq{defchi} we note that $r=\bP^1/\bP_2$, meaning that $r$ (and $\dot r$) has at most one cut on the main sheet $(-2h,2h)$.
At the same time $r=\bQ^\downarrow_1/\bQ^{2\downarrow}$, meaning that it has only one long cut at the same time. In other words the analytic continuation from above $r^\gamma$ is analytic in the LHP.
\beq\la{chi}
(\bQ^{k\downarrow})(u+i0)=r(u+i0) \epsilon^{kl}\bQ_l^{\downarrow}(u+i0)
=r(u+i0) \epsilon^{kl}G_{l}^{\;\;\dot n}\bQ_{\dot n}^\uparrow(u-i0),
\eeq
at the same time the l.h.s. can be expressed as
\beq
G^{k}_{\;\; \dot n}\;\bQ^{\dot n\uparrow}(u-i0)
=
G^{k}_{\;\; \dot n}\;\epsilon^{\dot n\dot m}\dot r(u-i0)\bQ_{\dot m}^{\uparrow}(u-i0)
=
\dot r(u-i0)
\epsilon^{kl}
G^{\;\; \dot n}_{l}
\bQ_{\dot n}^{\uparrow}(u-i0),
\eeq
from where we deduce that $ r(u+i0)=\dot r(u-i0)$.
Similarly, we can start from the dotted version of the derivation above to get $ r(u-i0)=\dot r(u+i0)$.
From this consideration we see that $ r(u)$ has a single quadratic cut, which connects it to $\dot r(u)$. This branch cut can be rationalised with the help of the Zhukovsky variable $x(u)=\frac{u+\sqrt{u-2h}\sqrt{u+2h}}{2h}$
so we can write $ r$
explicitly in terms of its zeros/poles\footnote{The number of poles and zeros $K_n$ is introduced to match later the notations in the ABA. $N_R$ and $N_B$ are introduced to allow for different types of Bethe roots to coincide and consequently cancel in the ratio.}
\beqa
r(u)=r_0\frac{\prod_i^{K_3-N_R} (x(u)-y_{3,i})
\prod_i^{K_{\dot 3}-N_R} (1/x(u)-y_{\dot 3,i})
}{
\prod_{i}^{K_1-N_B} (x(u)-y_{1,i})
\prod_i^{K_{\dot 1}-N_B} (1/x(u)-y_{\dot 1,i})
}\;,
\eeqa
and $\dot r(u)$ is $ r(u)$ with $x(u)$ replaced by $1/x(u)$.
$ r_0$ is a constant.
In the above expression
we assume $|y_{\dots }|\ge 1$.
Finding such a simple expression for a combination of $\bP$'s is an interesting novel feature of the $\AdS_3$ QSC.

\subsection{On analytic continuation}
\begin{figure}
    \centering
    \includegraphics[scale=0.6]{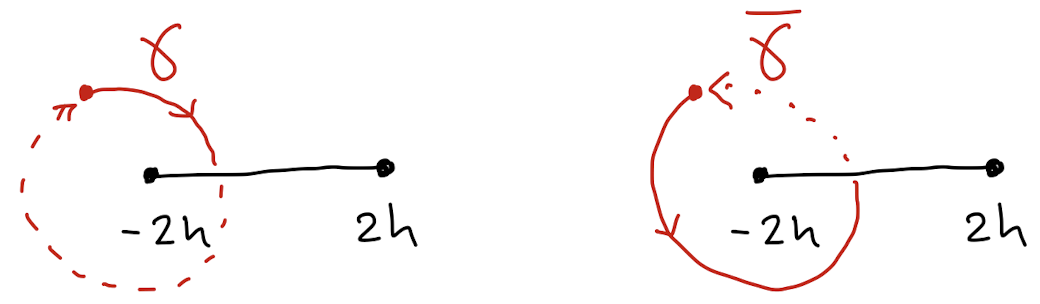}
    \caption{Two contours we  use for analytic continuation}
    \label{fig:gamma}
\end{figure}
We now deduce several consequences of the discussion in the previous section. We will see that the simple set of constraints given above implies the existence of a rich mathematical structure. The Q-functions live on a Riemann surface with infinitely many sheets, but the equations we will now deduce allow us to map any one of these sheets to the first one, as is the case also for the other examples of QSCs. 

As anticipated in the introduction, it will turn out that the branch cuts in this system of QSC equations \emph{cannot be} quadratic.  This means that, for any branch point on the Riemann surface, we can go around it in two ways, and in principle this yields two different results. 

We will introduce the analytic continuation paths $\gamma$ and its inverse $\gamma^{-1}$, which we will also denote by $\bar{\gamma}$. The  path $\gamma$ goes around a branch point at $2 h$ in anticlockwise sense, or alternatively, it goes around the branch point at $-2 h$ in clockwise sense.  
Since in this section we think in terms of short cuts for all the Q-functions, we can say that $\gamma$ goes through the short cut $(-2h, +2h)$ from above, while $\bar{\gamma}$ crosses it from below. The two paths are represented in figure~\ref{fig:gamma}. We denote the analytic continuation of any function of $u$ along these paths as $f^{\gamma}$ or $f^{\gamma^{-1}}\equiv f^{\bar{\gamma}}$. In this notation, (\ref{gluing1})
and (\ref{gluing2}) become
\beq\label{eq:Qtildebar2}
({\bQ}_k^\downarrow)^{\gamma} = G_k^{\;\; \dot m} \bQ^\uparrow_{\dot n} \;\;, \;\; ({\bQ}_{\dot k}^\downarrow)^{\gamma} = G_{\dot k}^{\;\; n} \bQ_{n}^\uparrow , \;\;\;\; ({\bQ}^{k\downarrow} )^{\gamma} = G^k_{\;\; \dot n} \bar{\bQ}^{\dot n\uparrow} , \;\;\;\; ({\bQ}^{\dot n \downarrow})^{\gamma} = G^{\dot a}_{\;\; b} \bQ^{ b\uparrow} .
\eeq

\subsubsection{The $\bQ\omega$-system}
By defining $i$-periodic functions $\omega$: 
\beq\label{eq:om-G-Om}
\omega_k^{\;\;\dot n} = G_k^{\;\;\dot m} \Omega_{\dot m}^{\;\; \dot n}
\,,\qquad\qquad \omega_{\dot k}^{\;\;  n} = G_{\dot k}^{\;\; m} \Omega_{  m}^{\;\;  n} \,,
\eeq
where $\Omega$'s are the matrices relating LHPA and UHPA bases in (\ref{eq:upOmega}), (\ref{eq:defOmega}),  the system of equations  (\ref{eq:Qtildebar2})  can be conveniently rewritten in the form
\beq\label{eq:Qomega}
\bQ^{\downarrow\gamma}_k = \omega_k^{\;\;\dot m} \bQ^\downarrow_{\dot m}
\,,\qquad\qquad\bQ_{\dot k}^{\downarrow\gamma} = \omega_{\dot k}^{\;\; m} \bQ^\downarrow_{ m} .
\eeq
Notice also that by construction, it follows from the properties of the gluing matrix and $\Omega$ function that  
\beq
\omega_k^{\;\;\dot m}  \omega^l_{\;\;\dot m} = \delta_k^l \,,\qquad\qquad\text{det}(\omega) = 1.
\eeq
Similarly, one can introduce
\beq
\bar\omega_{\dot n}^{\;\; m} = G^k_{\;\;\dot n}\Omega_{\;\; k}^{m} ,
\eeq
such that
\beq\label{eq:Qomega2}
\bQ^{\uparrow\gamma^{-1}}_{\dot k}= \bar\omega_{\dot k}^{\;\; m} \bQ^\uparrow_{ m} .
\eeq
In what follows, we adopt a simplified notation\footnote{Notice that this notation does not necessarily mean complex conjugation of the Q-functions; however, we expect that for real parameters there will be a simple relation. }, where
$\bQ^\downarrow$ is denoted by $\bQ$
and $\bQ^\uparrow$ is denoted by $\bar \bQ$. So \eq{eq:Qomega} and \eq{eq:Qomega2} become
\beq\label{eq:Qtil}
\bQ_k^\gamma = \omega_k^{\;\;\dot m}\bQ_{\dot m}\,,\qquad\qquad
\bar\bQ_k^{\bar\gamma} = \bar\omega_k^{\;\;\dot m}\bar\bQ_{\dot m}\;.
\eeq
Now let us understand the analytic continuation under the cuts of $\omega$, focusing on $\omega_{\dot k}^{\;\;l}$ first. Notice that the matrix $\Omega_k^{\;\;l}$ can be expressed
as $\Omega_k^{\;\;l} = \bar{Q}_{a|k}^+ Q^{a|l\;+}$ (see \eq{eq:defOmegaapp}) and since $Q^{a|i\; +}$ has no cut on the real axis, we only need to understand the analytic continuation of $\bar{Q}_{a|k}^+$.  The defining relation of this function is 
\beq
\bar Q_{a|k}^+ - \bar Q_{a|k}^- = \bP_a \bar \bQ_k ,
\eeq
where $\bar Q_{a|k}^-$ is now analytic and invariant under the analytic continuation along $\bar\gamma$.  Computing the discontinuity we obtain
\beq
\bar{Q}_{a|k}^{+\bar\gamma} - \bar{Q}_{a|k}^+ = \bP_a^{\bar\gamma} \bar{\bQ}_k^{\bar\gamma} - \bP_a \bar{\bQ}_k \; ,
\eeq
which, multiplied by
$Q^{a|l\;+}$ on the left, leads to
\beqa\label{eq:Omegatilde}
(\Omega_k^{\;\;l})^{\bar\gamma}-\Omega_k^{\;\;l}  =
\bar{\bQ}_k  ^{\bar\gamma} \bQ^{l\;\bar\gamma} - \bar{\bQ}_k \bQ^l\;. 
\eeqa
Next, multiplying these equations by $G_{\dot k}^{\;\;k}$, and using  (\ref{eq:Qtildebar2}), we find:
\beqa\label{eq:omegatil}
(\omega_{\dot k}^{\;\;l})^{\bar\gamma}-\omega_{\dot k}^{\;\;l}  =
{\bQ}_{\dot k}   \bQ^{l\;\bar\gamma} - 
{\bQ}^\gamma_{\dot k} \bQ^l\;.
\eeqa
This expression generalises a similar relation found in $AdS_5$ and $AdS_4$ cases, but now we distinguish two different directions for the analytic continuation on the r.h.s.. As usual one can replace dotted to undotted indices to get a similar identity for $\omega_{ k}^{\;\;\dot l}$. 

We can use \eq{eq:omegatil} to determine the double 
continuation of $\bQ_{\dot k}$ along the contour $\gamma$ -- we will then see explicitly that there may be an obstruction to the cuts being quadratic. 
We start by continuing (\ref{eq:Qtil}) along the inverse path $\bar{\gamma}$, which gives
\beq\la{derog}
\bQ_{\dot k}=(\omega_{\dot k}^{\;\;l})^{\bar\gamma}\bQ^{\bar\gamma}_l
=
\omega_{\dot k}^{\;\;l}\bQ^{\bar\gamma}_l - 
{\bQ}^\gamma_{\dot k} \bQ^l\bQ^{\bar\gamma}_l
=
\omega_{\dot k}^{\;\;l}\(\delta_l^p - 
{\bQ}_{l} \bQ^p\)\bQ^{\bar\gamma}_p ,
\eeq
where the second equality is obtained by using (\ref{eq:omegatil}), and recalling that $\bQ_i\bQ^i = 0$. 
Inverting the factor on the r.h.s., we get
\beq\label{eq:Qbargamma}
\bQ^{\bar\gamma}_p=
\(\delta_p^l + 
{\bQ}_{p} \bQ^l\)
\omega^{\dot k}_{\;\;l}\bQ_{\dot k}\;.
\eeq
From this we can compute directly the difference of the analytic continuation of the Q-function along $\gamma$ and $\bar{\gamma}$:
\beq
\bQ_k^{\gamma} - \bQ_k^{\bar{\gamma}} = \bQ_{\dot m} \left(\omega_k^{\;\;\dot m}-\omega^{\dot m}_{\;\;\; k} \right)+ \bQ_k \bQ^l \omega_l^{\;\;\dot m} \bQ_{\dot m}. 
\eeq
In the case of $AdS_5$, the two terms on the r.h.s. would vanish separately, due to the symmetry properties of the analogue of $\omega$, ensuring that the branch cuts are quadratic. In our case, that does not need to be the case,  since $\omega$ connects different kinds of indices and there is no reason \textit{a priori} to expect any symmetry between them. 

We make a further interesting observation by rewriting (\ref{eq:omegatil}) in the form
\beqa\label{eq:omegatil2}
(\omega_{\dot k}^{\;\;l})^{\bar\gamma}-{\bQ}_{\dot k}   \bQ^{l\;\bar\gamma}  =
\omega_{\dot k}^{\;\;l} 
-
{\bQ}^\gamma_{\dot k} \bQ^l\; \,.
\eeqa
This shows immediately that the combination $\omega_{\dot k}^{\;\;l} 
-
{\bQ}^\gamma_{\dot k} \bQ^l\;$ is equal to its analytic continuation, and therefore the cut on the real axis disappears in this combination. 
We can also write it as
$
\omega_{\dot k}^{\;\;m} 
(\delta_m^l
-
{\bQ}_{m} \bQ^l)
$.
Then taking \eq{derog} along $\gamma$, we get
\beq\label{eq:usefulgamma}
\bQ_{\dot k}^\gamma = 
\omega_{\dot k}^{\;\;l}\(\delta_l^p - 
{\bQ}_{l} \bQ^p\)\bQ_p =
\omega_{\dot k}^{\;\;l}\bQ_l 
\eeq
with the final equality being in agreement with~\eqref{eq:om-G-Om}. The first equality allows us to find the  expression for $\bQ_{\dot k}$ continued a second time along $\gamma$:
\beq\la{doubletrouble}
\bQ_{\dot k}^{\gamma^2} =
\omega_{\dot k}^{\;\;l}\(\delta_l^p - 
{\bQ}_{l} \bQ^p\)\bQ^\gamma_p =
\omega_{\dot k}^{\;\;l}\(\delta_l^p - 
{\bQ}_{l} \bQ^p\)\omega_p^{\;\;\dot h}\bQ_{\dot h}\,.
\eeq
This expression confirms the potential obstruction to the cuts being quadratic. 
In particular, we can repeatedly iterate this continuation and obtain in general
\beqa
&&(\bQ_{\dot k} )^{\gamma^n} =U_{\dot k}^{\;\; p}  (\bQ_p )^{\gamma^{n-1}} \,,\qquad\qquad(\bQ_{ k} )^{\gamma^n} = \dot U_{k}^{\;\; \dot p}  (\bQ_{\dot p} )^{\gamma^{n-1}}, \\
&&(\bQ_{\dot k} )^{\bar{\gamma}^n} =\bar{U}_{\dot k}^{\;\; p}  (\bQ_p )^{\bar{\gamma}^{n-1}} \,,\qquad\qquad(\bQ_{k} )^{\bar{\gamma}^n} = \dot{\bar{U}}_{k}^{\;\; \dot p}  (\bQ_{\dot p} )^{\bar{\gamma}^{n-1}},
\eeqa
where 
\beqa
&& U_{\dot k}^{\;\;p}\equiv \omega_{\dot k}^{\;\;l}\(\delta_l^p - 
{\bQ}_{l} \bQ^p\)\,,\qquad  \bar{U}_{\dot k}^{\;\;p}\equiv \left( \dot U^{-1} \right)_{\dot k}^{\;\;p} = \(\delta_{\dot k}^{\dot m} + 
{\bQ}_{\dot k} \bQ^{\dot m}\)
\omega^{p}_{\;\;\dot m} ,\\
&& \dot{U}_{ k}^{\;\;\dot p}\equiv \omega_{ k}^{\;\;\dot m}\(\delta_{\dot m}^{\dot p} - 
{\bQ}_{\dot m} \bQ^{\dot p}\), \;\;\;   \dot{\bar{U}}_{ k}^{\;\;\dot p}\equiv \left(  U^{-1} \right)_{ k}^{\;\;\dot p} = \(\delta_{  k}^{m } + 
{\bQ}_{ k} \bQ^{ m}\)
\omega^{\dot p}_{\;\; m} .
\eeqa
 In general, following the path $\gamma^n$ produces a concatenation of monodromies $U \cdot \dot U \cdot U \cdot \dot U \dots$, but since there is no reason to expect $(U \cdot \dot U )$ to be the identity matrix (or a  root of the latter), this is nontrivial, meaning that each branch point has infinite order and  connects to infinitely many sheets. 
 
Notice that, while in general we expect the branch points to be non-quadratic, there are some special combinations of Q-functions that do exhibit this property. We already showed that this is the case for the ratio $r$ defined in (\ref{defchi}). We now consider
\beq
\bQ^l\bQ^{\bar\gamma}_l=
\bQ^l
\omega^{\dot k}_{\;\;l}\bQ_{\dot k}=
\bQ^{\dot k\;\gamma}\bQ_{\dot k}\; ,
\eeq
where we used (\ref{eq:Qbargamma}) and the analogous equation to (\ref{eq:Qtil}) with (raised, dotted) indices. Lowering the indices with (\ref{defchi}), and remembering  that, as deduced above, $r^{\gamma} = r^{\bar{\gamma}} =\dot r$, the same relations (and their dotted version) can be written as
\beq\label{eq:aboveeq}
\epsilon^{kl}\bQ_k\bQ^{\bar\gamma}_l=
-\epsilon^{\dot k\dot l}\bQ_{\dot k}\bQ^{\gamma}_{\dot l}\,,\qquad\qquad
\epsilon^{kl}\bQ_k\bQ^{\gamma}_l=
-\epsilon^{\dot k\dot l}\bQ_{\dot k}\bQ^{\bar\gamma}_{\dot l}\;.
\eeq
Continuing the first equation above along $\gamma$, we get
\beq\label{eq:beloweq}
\epsilon^{kl}\bQ_k^\gamma\bQ_l=
-\epsilon^{\dot k\dot l}\bQ_{\dot k}^\gamma\bQ^{\gamma^2}_{\dot l} 
\; ,
\eeq
but due to the second equation in (\ref{eq:aboveeq}), the l.h.s. is also equal to $-\epsilon^{\dot k\dot l}\bQ_{\dot k}^{\bar\gamma}\bQ_{\dot l}$, meaning that the combination $\epsilon^{\dot k\dot l}\bQ_{\dot k}^{\bar\gamma}\bQ_{\dot l}=
\epsilon^{kl}\bQ_k\bQ^{\gamma}_l
$ comes back after $\gamma^2$!

As a final observation, we notice that, continuing the two sides of~\eqref{eq:omegatil2} along $\gamma$, one can also obtain an explicit equation for $\omega^{\gamma}$ in terms of quantities on the first sheet.

The main results of this section can be summarised in the following equations:\footnote{Results for $\bQ$ and $\omega$ functions with raised indices can be found using the same steps. }
\beq
(\bQ)^{\gamma}_{\dot k} = \omega_{\dot k}^{\;\; l} \bQ_l ,\;\; \;\;\;\; (\bQ^{\dot k} )^{\gamma} = \omega^{\dot k}_{\;\; l} \bQ^l ,\label{eq:Qomegafirst}
\eeq\,,\qquad\qquad
and
\beq
\left((\omega)^{\bar{\gamma}} - \omega\right)_{\dot k}^{\;\;  l}=\bQ_{\dot k}  (\bQ^{  l})^{\bar{\gamma}} -(\bQ_{\dot k})^{\gamma} \bQ^{ l} , \;\;\;\; \left((\omega)^{\bar{\gamma}} - \omega\right)^{\dot k}_{\;\; l}=-\bQ^{\dot k}  (\bQ_{ l})^{\bar{\gamma}} + (\bQ^{\dot k})^{\gamma} \bQ_{l}. \label{eq:Qomegalast} 
\eeq
Here, as usual, we understand that for every equation there is its double obtained by interchanging dotted and undotted indices. 
 Together with $\bQ_i\bQ^i = 0$, 
$
\omega_{\dot k}^{\; l} \omega^{\dot k}_{\; m }= \delta_m^l$, and the periodicity of $\omega$, the  relations (\ref{eq:Qomegafirst}),(\ref{eq:Qomegalast}) may be taken as a self-consistent description of the QSC, which is usually dubbed $\bQ\omega$-system.\footnote{As we saw in this section, these relations can be used to deduce algebraically all remaining properties, including the effect of crossing the cuts in the opposite directions. } 
  Bouncing back and forth between these equations, and using the fact that $\omega$ is $i$-periodic, one can obtain the result of any analytic continuation of the $\bQ$-functions and $\omega$ functions, inside any cut, and express it in terms of their values on the first sheet. This is the same feature that was observed in the other examples of QSC, see the discussion in \cite{Gromov:2013pga}. It is encouraging that this property is still valid here, even though the analytic structure is more complicated due to the branch points having infinite order. 

\subsubsection{The $\bP\mu$-system}
We now describe the constraints on the analytic continuation of $\bP$ functions. Analogously to~\cite{Gromov:2014caa}, the main object in this case is the matrix $\mu$ defined as
\beq\label{eq:defmu}
\mu_a^{\; \; \dot b} \equiv Q_{a|c}^- \, \omega^c_{\;\;\dot d} \,Q^{\dot b|\dot d \;\; -} \,,\qquad \mu^a_{\;\; \dot b} \equiv Q^{a|c\;-} \, \omega_c^{\;\;\dot d}\, Q_{\dot b|\dot d }^{ -} ,
\eeq
which will play a role similar to $\omega$. Notice that just like in the case of $\omega$, $\mu$ has unit determinant and $\mu_a^{\;\;\dot b} \mu^c_{\;\;\dot b} = \delta_a^c$.  We also notice the alternative expression
\beq\label{eq:barmu}
\mu_{a}^{\;\;\dot b} = \bar{Q}_{a|i}^- ( G_{\dot k}^{\;\; i} \omega^{\dot k}_{\;\; l} G^l_{\;\;\dot m} ) \bar{Q}^{\dot b| \dot m \; -} ,
\eeq 
which is obtained through the  relations (\ref{eq:QupQdown}), and will become useful in the discussion of the next section. 

\begin{figure}
    \centering
    \includegraphics[scale=0.5]{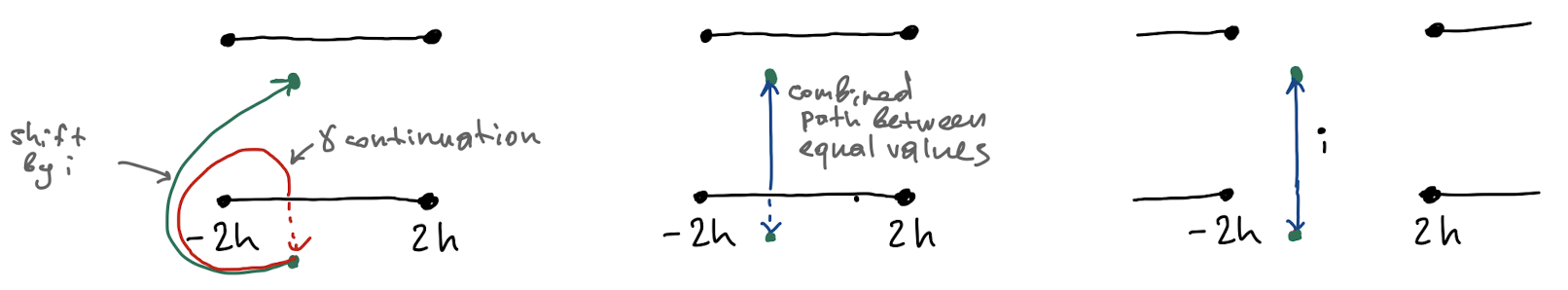}
    \caption{Periodicity of $\mu$ as a function with long cuts is identical to the property $\mu^{++}=\mu^\gamma$ for a section with short cuts.}
    \label{fig:mu}
\end{figure}

While $\omega$ is an $i$-periodic function on the Riemann section with short cuts, $\mu$  has a periodicity on the section with long cuts, as depicted in figure~\ref{fig:mu}.
Expressed in terms of a section with short cuts, this ``mirror periodicity'' becomes 
\beq\label{eq:mirrorperiodic}
\mu^{++} = (\mu)^{\gamma}.
\eeq
To prove this relation (and thus also long-cut periodicity of $\mu$), we continue it along $\bar{\gamma}$ and show that  the combination $(\mu^{++} )^{\bar{\gamma}} - \mu$ vanishes. We can rewrite such a difference as
\beqa
\left((\mu^{++} )^{\bar{\gamma}} - \mu \right)_a^{\;\;\dot b} = Q_{a|i}^+\;(\omega^{i }_{\;\dot k} )^{\bar{\gamma}}\; Q^{\dot b|\dot k \; +} - Q_{a|i}^- \;\omega^i_{\;\dot k} \; Q^{\dot b|\dot k \; -} .
\eeqa
We can now plug in $(\omega)^{\gamma^{-1}}$ from  the (undotted version of) (\ref{eq:Qomegalast}), and, in the second term, use the identities (\ref{eq:keytranslation}) to relate $Q_{a|i}^-$ and $Q_{a|i}^+$. We get
\beqa
&&Q_{a|k}^+\;\left( \omega^{k }_{\;\dot m} + ({\bQ}^k)^{\gamma} \bQ_{\dot m} -\bQ^k ( \bQ_{\dot m} )^{\bar{\gamma}} \right) Q^{\dot b |\dot m \; +} 
\nonumber \\
&& \qquad -\,\, Q_{a|k}^+ \left( \;\delta^k_m - \bQ^k \bQ_m \right) \omega^{m}_{\;\dot m} \left(\delta^{\dot m}_{\dot k} + \bQ^{\dot m} \bQ_{\dot k} \right)  Q^{\dot b|\dot k \; +} = 0 ,
\eeqa
where a perfect cancellation occurs due to (\ref{eq:Qomegafirst}),(\ref{eq:Qomegalast}), establishing  (\ref{eq:mirrorperiodic}).  
We can use $\mu$ to compute the values of $\bP$ on the second sheet. In particular, the definition (\ref{eq:defmu}), together with $\bP_a = Q_{a|i}^+ \bQ^i$, immediately implies
\beq\label{eq:Ptil0}
(\bP_a)^{\gamma} =\mu_a^{\;\dot b \;++} \bP_{\dot b} =(\mu_a^{\;\dot b })^{\gamma} \bP_{\dot b},
\eeq
which is conveniently rewritten as
\beq\label{eq:Ptil}
(\bP_a )^{\bar{\gamma}} = \bP_{\dot b} \mu^{\dot b}_{\;\; a} \;. 
\eeq
This equation, compared to (\ref{eq:Qtil}), highlights the symmetry of the construction between $\bP$ and $\bQ$ functions. 
From (\ref{eq:omegatil}), it is also immediate to derive
\beq
(\mu_a^{\;\;\dot b})^{\gamma} - \mu_a^{\;\;\dot b} = Q_{a|k}^+\left( \omega^k_{\;\;\dot m} - (\omega^k_{\;\;\dot m})^{\bar{\gamma}} \right) Q^{\dot b|\dot m+} = \bP_a (\bP^{\dot b} )^{\bar{\gamma}} - (\bP_a)^\gamma \bP^{\dot b},
\eeq
which shows that the combination $\mu_a^{\;\;\dot b} + \bP_a (\bP^{\dot b} )^{\bar{\gamma}} = \left(\delta_a^b + \bP_a \bP^b\right)\mu_b^{\;\;\dot b}$ has no cut on the real axis. From this observation and (\ref{eq:Ptil}) we also deduce
\beq
(\bP_a)^{\gamma} = \left(\delta_a^b + \bP_a \bP^b\right)\mu_b^{\;\;\dot b}\bP_{\dot b}\,,\qquad\qquad (\bP_a)^{\gamma^2} = \left(\delta_a^b + \bP_a \bP^b\right)\mu_b^{\;\;\dot b}(\bP_{\dot b})^{\gamma} ,
\eeq
and we obtain, similar to the previous discussion, that the branch points are in general connected to an infinite series of sheets, which can be reached by iterating
\beqa
(\bP_a )^{\bar{\gamma}^n} = \bar{W}_{a}^{\;\;\dot b}( \bP_{\dot b} )^{\bar{\gamma}^{n-1} } \,,\qquad\qquad (\bP_a )^{{\gamma}^n} = W^{\dot b}_{\;\;a}( \bP_{\dot b} )^{{\gamma}^{n-1} }  ,
\eeqa
with $W$, $\bar{W}$ defined by
\beq
W_a^{\;\;\dot b} = \left(\delta_a^b + \bP_a \bP^b\right)\mu_b^{\;\;\dot b}\,,\qquad\qquad \bar{W}_a^{\;\;\dot b} = \mu^{\dot c}_{\;\;a}\left(\delta_{\dot c}^{\dot b} - \bP_{\dot c} \bP^{\dot b}\right),
\eeq
with $\dot{ W } = ( \bar{W} )^{-1}$, $\dot{\bar{W}}= (W )^{-1}$ defined similarly by dotting/undotting all indices. As in the previous paragraph, we see that going around the branch point many times keeps leading to new sheets, since we expect in general that  $(W \cdot \dot W )_a^{\;b}\neq \delta_a^b$, being there no reason to expect otherwise. 

We can summarise the finding of this section in a set of $\bP\mu$ equations. For the first wing they read,
\beq
(\bP_a)^{\bar{\gamma}}  = \bP_{\dot b} \mu^{\dot b }_{\;\; a}\,,\qquad\qquad (\bP^a)^{\bar{\gamma}}  = \bP^{\dot b} \mu_{\dot b }^{\;\; a} ,\label{eq:Pmufirst}
\eeq
and
\beq
(\mu_a^{\;\;\dot b})^{\gamma} - \mu_a^{\;\;\dot b} =  \bP_a (\bP^{\dot b} )^{\bar{\gamma}} - (\bP_a)^\gamma \bP^{\dot b}, \;\;\;\; (\mu^a_{\;\;\dot b})^{\gamma} - \mu^a_{\;\;\dot b} =  -\bP^a (\bP_{\dot b} )^{\bar{\gamma}} + (\bP^a)^\gamma \bP_{\dot b}. \label{eq:Pmulast}
\eeq
 Together with the mirror-periodicity of $\mu$, this can also be taken as a self-consistent description of the QSC. As  remarked for the $\bQ\omega$-system, these equations contain enough information to map the values of $\bP$ and $\mu$ functions on any sheet, back to the first main one. 
 
\section{The ABA limit}\label{sec:ABAQSC}
In this section, we will find an asymptotic solution for some of the Q-functions in the large-$L$ limit. This will lead us to a perfect match with the Asymptotic Bethe Ansatz for massive states, including the dressing phases. 

\subsection{Large-volume scaling of the QSC}
To deduce the large-$L$ solution, we will use arguments developed for the $AdS_5$ case in \cite{Gromov:2014caa}
and then also successfully used for $AdS_4$
case to derive the ABA in \cite{Bombardelli:2017vhk}. The crucial observation is that, for large $L$, some Q-functions are exponentially suppressed/enhanced, following the pattern of their large-$u$ asymptotics (\ref{eq:largeu}). 
Following the notation of \cite{Gromov:2014caa}, we introduce a parameter $\epsilon \propto e^{-L}$ to keep track of this scaling. 
We then see that for large $L$ (i.e., $\epsilon \sim 0$),
\beqa\label{eq:epsilonQ}
&&Q_{a|i} \sim \left(\begin{array}{cc} 1  & \epsilon^2\\
\frac{1}{\epsilon^2} & 1
\end{array} \right) , \;\;\;\; Q^{a|i} \sim \left(\begin{array}{cc} 1 & \frac{1}{\epsilon^2}\\
{\epsilon^2} & 1
\end{array} \right) ,\\
&&\bQ_i \sim (\epsilon^{-1} , \epsilon ), \; \bQ^i \sim (\epsilon, \epsilon^{-1} ) , \;\;\; \bP_i \sim (\epsilon , \epsilon^{-1} ), \; \bP^i \sim (\epsilon^{-1}, \epsilon ).\label{eq:epsilonQ2}
\eeqa
In the second wing, we would have exactly the same pattern for the dotted Q-functions. In addition, since the $\omega$ functions are periodic on a Riemann section with short cuts, they have constant asymptotics. We will then assume that they all scale as
\beq
\omega_k^{\;\dot m}\sim O(1) ,\; \omega^k_{\;m} \sim O(1),\; \omega_{\dot k}^{ \;m}\sim O(1) , \;  \omega^{\dot k}_{\;m}\sim O(1), \;\;\;  \epsilon\rightarrow 0.
\eeq
We then notice that some of the QQ relations, $\bP\mu$ and $\bQ\omega$ equations simplify significantly. 
Dropping the subleading terms for $\epsilon\rightarrow 0$ we find for instance, from (\ref{eq:defmu}),
 \beq
 \mu_1^{\;\dot 2} = Q_{1|k}^- \omega^k_{\;\; \dot l} Q^{ \dot 2 | \dot l \; -} \sim Q_{1|1}^- \omega^1_{\;\; \dot 2} Q^{ \dot 2 | \dot 2 \; -} = Q_{1|1}^- \omega^1_{\;\; \dot 2} Q_{ \dot 1 | \dot 1}^{ -} ,
 \eeq
 and similarly we get to
 \beq\label{eq:muABAdef}
 \mu^{2}_{\;\;\dot 1} \sim Q_{1|1}^- \omega_2^{\;\; \dot 1} Q_{ \dot 1 | \dot 1}^{ -},
 \eeq
 where we recalled that by definition $Q^{2|2} = Q_{1|1}$. 
 Another important equation is obtained starting from $\bP_1 = (Q_{1|i}^+) \bQ^i$, and considering the analytic continuation along $\gamma$ (recall that $Q_{a|1}^+$ has no cut on the real axis). Using the $\bQ\omega$-system, and then considering the large-$L$ scaling,  we get
 \beq\label{eq:Ptilde}
 ({\bP}_1 )^{\gamma} = Q_{1|l}^+ \omega^{l}_{\; \dot k} \bQ^{\dot k} \sim Q_{1|1}^+ \omega^{1}_{\; \dot 2} \bQ^{\dot 2} , \eeq
 which will play an important role in the following derivation of the ABA. 

We now proceed to deduce the form of some of the elements of the QSC in the ABA scaling. 
To do that, we will take as a working hypothesis the property that, for the functions $\mu_1^{\;\;\dot 2}$, $\mu_{\dot 1}^{\;\;2}$, $\mu^{\dot1}_{\;\;2}$, $\mu^{1}_{\;\;\dot 2}$, the cut on the real axis becomes quadratic in the large-$L$ limit. 
We will see that all the solutions for massive states fall into this category.\footnote{It is tempting to speculate that asymptotic solutions including massless modes might be found by relaxing this assumption on the behaviour at large $L$. On the other hand the massless modes suffer from stronger wrapping effects, which limits the range of validity of the corresponding ABA regime, which may mean that the approach of \cite{Gromov:2014caa} is not   sensitive enough to detect those power-like effects, and the ABA should be recovered via a different route. We reserve these questions for future studies.}

We will also make an assumption that the gluing matrix follows the  pattern one can deduce from the  gluing equations in the classical limit, namely that all the diagonal elements vanish.
Our derivation assumes that this is true at least in the ABA limit, but we suspect it may be true even at finite $L$ (this is what happens in $AdS_5$). 

Finally, we will use the expressions obtained from (\ref{eq:barmu}) in the ABA limit, such as
 \beq\label{eq:muABAbar}
\mu_{1}^{\;\; \dot 2} \sim \bar{Q}_{1|1}^- ( G_{\dot k}^{\;\; 1} \omega^{\dot k}_{\;\; l}  G^l_{\;\;\dot 2} ) \bar{Q}_{\dot 1|\dot1}^- \propto \bar{Q}_{1|1}^- \omega^{\dot 2}_{\;\; 1}   \bar{Q}_{\dot 1|\dot1}^- .
 \eeq

\subsection{Fixing Q-functions on the first sheet}
\paragraph{Finding $Q_{1|1}$, $\mu_1^{\;\; \dot 2}$ and $\omega^1_{\;\;\dot 2}$. }
To determine these functions, we use the assumption on the quadratic nature of the branch point of $\mu_1^{\;\;\dot 2}$ in the ABA limit. Even though this assumption could appear to be too restrictive, we will nevertheless show that in the ABA limit this extra restriction does not lead to any inconsistencies. The simplification of the analytic structure of $\mu$ is quite typical  in the ABA limit -- for instance in the $AdS_5$ case the discontinuity of $\log\mu$ appears to be a simple rational function of $x$, whereas in general it would have an infinity tower of cuts.
With that in mind, we can follow closely \cite{Gromov:2014caa}, and this part may be  skimmed through by the reader familiar with that paper. The surprises begin from section \ref{sec:surprises}, where the non-quadratic nature of the branch points pops up again in a crucial way.

We start by considering the function $\mu_1^{\;\;\dot 2}(u+i/2)$. We take it to have a finite number of zeros on the first Riemann sheet with short cuts, and we store such zeros in a polynomial $\mathbb{Q}(u)=\prod_i(u-u_i)$. We then consider
\beq
(F)^2 \equiv \frac{\mu_1^{\;\; \dot 2} }{\mu_1^{\;\; \dot 2 ++} } \frac{\mathbb{Q}^+}{\mathbb{Q}^-} = \frac{\mu_1^{\;\; \dot 2} }{(\mu_1^{\;\; \dot 2})^{\gamma} } \frac{\mathbb{Q}^+}{\mathbb{Q}^-}
\eeq
which by definition has no zeros or poles on the first Riemann sheet with short cuts. Since by our assumption the branch points of $\mu$ become quadratic in the ABA limit, using the property $(\mu)^{\gamma} = \mu^{++} \sim \mu^{\bar\gamma}$, it is simple to obtain the same equations as in \cite{Gromov:2014caa}:
\beq\label{eq:FRH}
F (F)^{\gamma} \sim F (F)^{\bar\gamma} \sim \frac{\mathbb{Q}^+}{\mathbb{Q}^-}.
\eeq
All the other cuts in $F$ must disappear in the ABA limit. In fact, using (\ref{eq:muABAdef}), and the periodicity of $\omega$, we see that $F^2$ can be rewritten as
\beq\label{eq:Q11mu}
 F^2 =\frac{Q_{1|1}^- \, Q_{\dot 1|\dot 1}^-}{Q_{1|1}^+ \, Q_{\dot 1|\dot 1}^+}\frac{\mathbb{Q}^+}{\mathbb{Q}^-} ,
\eeq
which does not have cuts in the upper half plane, while (\ref{eq:muABAbar}) leads us to the expression 
\beq
 F^2 = \frac{\bar{Q}_{1|1}^- \, \bar{Q}_{\dot 1|\dot 1}^-}{\bar{Q}_{1|1}^+ \, \bar{Q}_{\dot 1|\dot 1}^+}\frac{\mathbb{Q}^+}{\mathbb{Q}^-} ,
\eeq
which shows that there are no cuts in the lower half plane either.
Taking into account that $F$ has constant asymptotics at large $u$ on the first sheet, we have a simple  Riemann-Hilbert problem (\ref{eq:FRH}), with the standard solution
 \beq
 F = \pm e^{i \frac{\mathcal{P}}{2} } \frac{B_{(+)}}{B_{(-)}},
 \eeq
 with $e^{i\mathcal{P}}\equiv\prod_{i} \frac{x_i^+}{x_i^-}$,
 and 
 $B_{(\pm)}(u)\equiv \sqrt{\frac{h}{x_i^{\mp}}} (\frac{1}{x(u)} - x_i^{\mp} )$.  The constant factor will not be very important in the current considerations.\footnote{In any case, one can establish by an argument parallel to the one in \cite{Gromov:2014caa}, that $e^{i\mathcal{P}} =1$, which can be recognised as the level matching condition in the  ABA interpretation.}

Setting $Q_{1|1} Q_{\dot 1|\dot 1} \equiv \mathbb{Q} (f^+)^2$, equation (\ref{eq:Q11mu}) then gives us a difference equation
\beq
\frac{f^{++}}{f} = \frac{B_{(-)}}{B_{(+)}} ,
\eeq
where by construction $f$ should have neither poles nor zeros in the upper half plane, and power-like asymptotics. 
Up to a multiplicative constant, the solution is
\beq\label{eq:intf}
f(u) \propto \text{exp} \left(\int_{-2h}^{2h}\frac{dz}{2 \pi i}\log\frac{B_{(-)}(z + i 0^+)R_{(+)}(z+ i 0^+)}{B_{(+)}(z+ i 0^+)R_{(-)}(z+ i 0^+)} \partial_z \log\Gamma( i z-iu ) \right), 
\eeq
where we use $\propto$ to indicate that there could be an irrelevant constant factor in the equation.
With the explicit form of $f$ in \eq{eq:intf}, we have fixed   $Q_{1|1} Q_{\dot 1|\dot 1}$ completely. 
Noticing that $\mu_1^{\;\;\dot 2} = Q_{1|1}^- Q_{\dot 1|\dot 1}^-\omega^1_{\;\;\dot 2} \propto \bar{Q}_{1|1}^- \bar{Q}_{\dot 1|\dot 1}^- \omega^1_{\;\;\dot 2}$, where $\omega$ should be $i$-periodic,  we can also find
\beq
\mu_1^{\;\;\dot 2} \propto\mathbb{Q}^- f \bar{f}^{--}, \;\;\; \omega^1_{\;\;\dot 2} \propto \frac{\bar{f}^{--} }{f} , \;\;\; Q_{1|1} Q_{\dot 1| \dot 1} \propto \mathbb{Q} \, ( f^+ )^2 ,
\eeq
where  
 $\bar{f}$ is solution of $\bar{f}/\bar{f}^{--} = \frac{B_{(-)}}{B_{(+)}}$ with no cuts in the lower half plane and constant asymptotics.\footnote{ We have that $\bar{f}$ is simply the complex conjugate of $f$ for real roots, and otherwise it is given by a simple integral representation similar to (\ref{eq:intf}).}  From the expression (\ref{eq:Q11mu}), we also see that the set of zeros of $\mathbb{Q}$ must coincide with the union of the zeros of $Q_{1|1}$ and $Q_{\dot 1|\dot 1}$. Therefore we split this polynomial as
 $
\mathbb{Q}(u) \equiv \mathbb{Q}_2(u) \mathbb{Q}_{\dot 2}(u) $, with $ \mathbb{Q}_2(u)\equiv \prod_{i=1}^{K_2}(u-u_{2,i})$, $ \mathbb{Q}_{\dot 2}(u)\equiv  \prod_{i=1}^{K_{\dot 2}}(u-u_{\dot 2,i})$, 
 with the understanding that $\mathbb{Q}_2$ contains zeros of $Q_{1|1}$, and $\mathbb{Q}_{\dot 2} $ zeros of $Q_{\dot1|\dot1}$. This notation is chosen in anticipation of the role of the zeros in the ABA. 
With these conventions, we have
\beq\label{eq:expl1}
\mu_1^{\;\;\dot 2} \propto \mathbb{Q}^-_2 \mathbb{Q}^-_{\dot 2}  f_2 f_{\dot 2} \bar{f}^{--}_2  \bar{f}^{--}_{\dot 2}, \;\;\; \omega^1_{\;\;\dot 2 } \propto  \frac{\bar{f}^{--}_2  \bar{f}^{--}_{\dot 2} }{f_2 \, f_{\dot 2}} , \;\;\; Q_{1|1}  \propto \mathbb{Q}_2  \,  f^+_2 f^+_{\dot 2}\, P, \;\;\;  Q_{\dot 1|\dot 1}  \propto  \mathbb{Q}_{\dot 2} \,  f^+_2 f^+_{\dot 2} \frac{1}{P},
\eeq
with the obvious notation that $f_{\alpha}$ are solutions of $f_{\alpha}^{++}/f_{\alpha}= \frac{B_{\alpha,(-)}}{B_{\alpha,(+)}}$, with $\alpha \in \left\{2,\dot 2\right\}$ (see appendix \ref{app:ABA}), and where $P$ is a yet unfixed function of $u$ coming from splitting the product $Q_{1|1} Q_{\dot 1|\dot 1}$. This function should have neither zeros nor poles, and moreover $P^-$ cannot have any cuts in the upper half plane. On the other hand, the quantity
\beq\label{eq:LHPA4}
\bar{Q}_{1|1}^+ = Q_{1|j}^+ \Omega_1^{\;\;j} \sim Q_{1|1}^+ \Omega_1^{\;\;1} \;\;\;\text{ for  } \epsilon\rightarrow 0,
\eeq
should be analytic in the lower half plane, where the matrix $\Omega$ is defined by $\omega_{\dot a}^{\;\; b} = G_{\dot a}^{\;\; c} \, \Omega_{ c}^{\;\; b}$. 
 Using the assumed classics-inspired off-diagonal property of the gluing matrix, we see that $\Omega_1^{\;\;1} \propto \omega_{\dot 2}^{\;\; 1}$. 
Then, from (\ref{eq:LHPA4}) and the above found solution for $\omega_{\dot 2}^{\;\; 1}$, we deduce that $\bar{Q}_{1|1}^+$ - which should be analytic in the lower half plane - can also be written as 
 $
\mathbb{Q}_2 \frac{B_{2,(-)} B_{\dot 2,(-)}}{B_{2,(+)} B_{\dot 2,(+)}} \bar{f}_2^{--}\, \bar{f}_{\dot 2}^{--}\, P^+ .
$
Since all the other factors already have this property, we conclude that $P^+$ should have no cuts in the lower half plane. 
All together, we found that the function $P$ cannot have any singularities or zeroes and thus is a constant (due to regularity at infinity).  
In conclusion, we found
\beqa
&& Q_{1|1} \propto \mathbb{Q}_2 \, f_{\dot 2}^+ f_2^+ , \;\;\; Q_{\dot 1|\dot 1} \propto \mathbb{Q}_{\dot 2} \, f_{\dot 2}^+ f_2^+ ,\label{eq:Q11fixed}
\\
\nn&&\mu_1^{\;\;\dot 2} \propto \mu_{\dot 1}^{\;\; 2} \propto 
\mathbb{Q}_2^- \, \mathbb{Q}_{\dot 2}^- \;f_2 \, \bar{f}_{ 2}^{--}   f_{\dot 2} \, \bar{f}_{ \dot 2}^{--} ,
\\
\nn&&
\omega^1_{\;\;\dot 2} \propto \omega^{\dot 1}_{\;\; 2} \propto 
 \frac{\bar{f}_{ 2}^{--} }{f_2} \frac{ \bar{f}_{ \dot 2}^{--} }{f_{\dot 2} } ,\label{eq:omegas} 
\eeqa
where we included the values of more $\omega$, $\mu$ functions, obtained by obvious generalisations of the argument above. 

\paragraph{Parametrising $\bP$ and $\bQ$ functions. }
From the $AdS_5$ and $AdS_4$ cases, we expect that a special subset of $\bP$ and $\bQ$ functions will converge to simple explicit expressions in the ABA limit. This is the subset of the $\bP$ functions which are \emph{small}, together with the $\bQ$ functions that are \emph{large}, for $\epsilon\rightarrow 0$. From (\ref{eq:epsilonQ2}), we see that those are $\bP_1$, $\bP^2$, $\bQ_1$, $\bQ^2$, and their dotted counterparts. We expect that their zeros on the first sheet will acquire the meaning of Bethe roots. 

With this in mind, we make the following ansatz:
\begin{align}\label{eq:Qide}
\bP_1 &\propto x^{-L/2} \,\mathcal{A} \times R_{\tilde{1} } B_{\tilde{\dot  1}} \,
B_{2, (-)}  , & 
\bP^2 &\propto x^{-L/2} \,\mathcal{A} \times R_{\tilde 3} B_{\tilde{\dot 3}} B_{2, (-)  } ,\\
\bQ_1  &\propto \frac{x^{L/2} }{\mathcal{A}' } \times R_{1} B_{\dot 1}
f_{2}\frac{f_{\dot 2} }{B_{\dot 2, (+) } } , &
\bQ^2 &\propto  \frac{x^{{L}/2} }{\mathcal{A}'} \times R_{3} B_{\dot 3} \, f_{2}  \frac{f_{\dot 2} }{B_{\dot 2, (+) } }\;.\label{eq:Qidelast}
\end{align}
Above, we have stored the zeros of the $\bP$ and $\bQ$ functions on the first sheet inside the Zhukovsky polynomials $R_{\alpha}$, defined\footnote{In the definitions (\ref{eq:defR}),(\ref{eq:defB}), we take the zeros to satisfy $|x_{\alpha, j}|>1$, which means the zeros of $R_{\alpha}$ ($B_{\alpha}$) are on the first (second) sheet in terms of the spectral parameter $u$. } in appendix \ref{app:ABA} (again, the notation anticipates the role of these zeros in the ABA, but for now they are generic parameters). 
The other $B_{\alpha}$ and $f_{\alpha}$  factors (also defined in the appendix) are chosen for future convenience, but they do not have zeros on the first sheet. 
Notice that the ansatz above is fully general, because it contains the  arbitrary functions of $u$ $\mathcal{A}(u)$, $\mathcal{A}'(u)$. By construction they should have no poles or zeros on the first sheet, and moreover $\mathcal{A}$, which appears in the $\bP$ functions, can have only a single cut. 

Comparing with  \eq{defchi}, we see that we can write the important function $r$ in two alternative ways as
\beq
r\propto \frac{R_{\tilde 1}B_{\tilde {\dot 1}}}{R_{\tilde 3}B_{\tilde{\dot 3}}}\propto
\frac{R_3 B_{\dot 3}}{R_1 B_{\dot 1}}\;,
\eeq
which means that $R_{\tilde 3}$ and $R_{\tilde 1}$ could have common zeroes.

Likewise we make a similar ansatz for the second wing:
\begin{align}
\bP_{\dot 1} &\propto x^{-{L}/2} \,\dot{\mathcal{A}}  \times R_{\dot 3}' B_{3}'  \, B_{\dot 2,(-)} , & \bP^{\dot 2} &\propto  x^{-{L}/2} \,\dot{\mathcal{A}} \times R_{\dot 1}' B_{1}' \, B_{\dot 2, (-)  }, \label{eq:Qsnd}\\
\bQ_{\dot 1}  &\propto \frac{x^{{L}/2} }{\dot{\mathcal{A}}' } \times R_{\tilde{\dot  3} }' B_{\tilde 3 }' f_{\dot 2}  \, \frac{f_{2} }{B_{2, (+) } } ,  &
\bQ^{\dot 2} &\propto \frac{x^{{L}/2} }{\dot{\mathcal{A}}' } \times R_{\tilde{\dot 1}}' B_{\tilde 1}' \, f_{\dot 2}  \, \frac{f_{2} }{B_{2,(+) } }\;,\label{eq:Qsndlast}
\end{align}
with functions $\dot{\mathcal{A}}(u)$, $\dot{\mathcal{A}}'(u)$ having no zeroes on the main sheet. In (\ref{eq:Qsnd}),(\ref{eq:Qsndlast}), we have introduced polynomials in $x$ vs $\frac{1}{x}$, $R_{\alpha}'$ and $B_{\alpha}'$, respectively. They are defined just like in (\ref{eq:defB}),(\ref{eq:defR}), but where the zeros of these polynomials (and their number) are in principle unrelated to the ones appearing in the first wing. We will however soon see that there is a simple identification. 
From \eq{defchi} we again get
\beq
\dot r \propto \frac{R'_{\dot 3} B'_3}{R'_{\dot 1} B'_1}\propto
\frac{R'_{\tilde{\dot 1}} B'_{\tilde 1}}{R'_{\tilde{\dot 3}} B'_{\tilde 3}}\;.
\eeq
Furthermore, recalling that $\dot r^\gamma=r$ we get
\beq\la{RBall}
\frac{R_{\tilde 1}B_{\tilde {\dot 1}}}{R_{\tilde 3}B_{\tilde{\dot 3}}}\propto
\frac{R_3 B_{\dot 3}}{R_1 B_{\dot 1}}
\propto \frac{B'_{\dot 3} R'_3}{B'_{\dot 1} R'_1}\propto
\frac{B'_{\tilde{\dot 1}} R'_{\tilde 1}}{B'_{\tilde{\dot 3}} R'_{\tilde 3}}\;.
\eeq
One can for example deduce that $R_{\tilde 1} R_1=R_3R_{\tilde 3}$ etc. from the above equation.

\paragraph{Fermionic duality equation. }
An important constraint comes from one of the QQ relations
\beq\label{eq:QQtwoeq}
Q_{1|1}^+-Q_{1|1}^- = \bQ_1 \bP_1\;,
\eeq
where we see the appearance of $Q_{1|1}$ determined in (\ref{eq:Q11fixed}). Plugging in that value, and the ansatz (\ref{eq:Qide}),(\ref{eq:Qidelast}), we find, from the first equality in (\ref{eq:QQtwoeq}),
\beq\la{RBall2}
R_{2,(+)} B_{\dot 2, (-)}-R_{2, (-)} B_{\dot 2, (+)} \propto R_1 \, R_{\tilde 1} B_{\dot 1} B_{\tilde{ \dot 1}} \frac{\mathcal{A}}{\mathcal{A}'}\; , 
\eeq
where we used the property that $\mathbb{Q}_{\alpha}^{\pm} =B_{\alpha, (\pm)}R_{\alpha, (\pm)}$. 
Since the left hand side is a rational function in $1/x(u)$, and $\mathcal{A}$, $\mathcal{A}'$ should have no zeros or poles on the first sheet, the ratio $\mathcal{A}(u)/\mathcal{A}'(u)$ can only  be a polynomial in the variable $\frac{1}{x(u)}$. But we can absorb any such function in a redefinition of the $B_{\tilde{\dot 1}}$, $B_{\dot 1}$ polynomials (which are so far completely unconstrained), so without loss of generality we can take $\mathcal{A}/\mathcal{A}' = 1$. 
Similar considerations arise from considering (\ref{eq:QQtwoeq}) in the second wing. From now on, therefore we take
\beq
\mathcal{A}(u) = \mathcal{A}'(u) \;\;,\;\; \dot{\mathcal{A}}(u) = {\dot{\mathcal{A}}}'(u).
\eeq
Notice that we still have two undetermined functions, which will be fixed in the next section. 
Using that ${\cal A}={\cal A}'$, 
from (\ref{RBall2}) we obtain
\beq\label{eq:fermdual1}
R_{2,(+)} B_{\dot 2, (-)}-R_{2, (-)} B_{\dot 2, (+)} \propto R_1 \, R_{\tilde 1} B_{\dot 1} B_{\tilde{ \dot 1}}\;.
\eeq
The analogous constraint obtained by considering the second wing reads
\beq\label{eq:fermdual2}
B_{2, (-)} R_{\dot 2, (+)}-B_{2,(+)} R_{\dot 2, (+)} \propto B_1' \, B_{\tilde 1}' R_{\dot 1}' R_{\tilde{ \dot 1}}'\;,
\eeq
and analytically continuing this equation to another sheet we find the identity
\beq
R_1 \, R_{\tilde 1} B_{\dot 1} B_{\tilde{ \dot 1}} \propto R_1' \, R_{\tilde 1}' B_{\dot 1}' B_{\tilde{ \dot 1}}'
\eeq
which implies
\beq
R_1 \, R_{\tilde 1}=R_1' \, R_{\tilde 1}'\;.
\eeq

Equations of the form (\ref{eq:fermdual1}) are examples of fermionic duality relations. They imply that the sets of roots with labels $1$, $2$, $3$ (or alternatively the ``dual'' set obtained with $1\leftrightarrow \tilde{1}$, $3\leftrightarrow \tilde{3}$) satisfy the auxiliary ABA equations of the form
\beq\label{eq:BAauxiliary}
1 =\left. \frac{\mathbb{Q}^+_2 B_{2,(-)}\,B_{\dot 2,(-)}}{\mathbb{Q}^-_2 B_{2,(+)}\,B_{\dot 2,(+)}} \right|_{u \in \left\{\text{roots of type } 1,\tilde{1}, 3,\tilde{3}\right\}}.
\eeq

\subsection{Going inside the cut: fixing the dressing phases}\label{sec:surprises}
So far we reduced the ansatz for $\bP$'s and $\bQ$'s to just two unknown functions with one cut and no zeroes ${\cal A}$ and $\dot{\cal A}$
on the main sheet. In order to constrain them further, we need to go to the next sheet of their Riemann surfaces.

This will bring us to the most interesting part of the analysis, where things will be radically different than in $AdS_5$ and $AdS_4$. 
By studying equations of the form (\ref{eq:Ptilde}),  which we repeat here,
 \beqa\label{eq:P1tildeABA2}
&& ({\bP}_{1})^{\gamma} \sim  Q_{1| 1}^+ \omega^{ 1}_{\;  \dot 2} \bQ^{\dot 2}
\;\;,\;\; ({\bP}_{\dot 1})^{\gamma} \sim  Q_{\dot 1| \dot 1}^+ \omega^{ \dot 1}_{\;  2} \bQ^{ 2} 
\label{eq:P1tildeABA3}
\eeqa
we will find that the $\bP$ and $\bQ$ functions \emph{cannot have} a quadratic cut even in the ABA limit. We will also be able to fix the form of the yet undetermined functions $\mathcal{A}(u)$, $\dot{\mathcal{A}}(u)$ and relate them to the dressing phases of~\cite{Borsato:2013hoa}.

\subsubsection{The cuts cannot be quadratic}\label{sec:notquadratic}
The strategy will be to compare  the r.h.s. of each of the equations (\ref{eq:P1tildeABA2}), with the analytic continuation of $\bP$ functions, starting from their form in (\ref{eq:Qide}),(\ref{eq:Qsndlast}).\footnote{
Here a comment is in order: in principle, the analytic continuation through the cut might not commute with the large-$L$ limit, due to the presence of Stokes-type phenomena - where a subleading correction on the first sheet might become large on the second sheet invalidating the result. However, as discussed in \cite{Gromov:2014caa}, one can expect that it is safe to analytically continue the ABA limit of a Q-function that is already \emph{small} on the first sheet. This is the case of the $\bP$ functions we consider which are of order $\epsilon$. }
 From the first equation in (\ref{eq:Ptilde}), in particular, we obtain:
\beq
\bP_1^\gamma=x^{L/2}({\mathcal{A}} )^{\gamma}
B_{\tilde{1} } R_{\tilde{\dot  1}} \,
R_{2, (-)}=
\({\mathbb Q}_2^+ 
f_{\dot 2}^{++}
f_{2}^{++}
\)
\(
\frac{\bar f_2\bar f_{\dot 2}}{
f_2^{++}
f_{\dot 2}^{++}
}
\)
\(
\frac{x^{L/2}}{\dot{\cal A}}
R_{\tilde {\dot 1}}'
B'_{\tilde 1}
\frac{f_2 f_{\dot 2}}{B_{2,(+)}}
\)\;.
\eeq
We noticed in the previous section that the roots of $R_{\tilde 1}$ and 
$R'_{\tilde 1}$ satisfy the same BAE
equation \eq{eq:BAauxiliary}. The same is true for the roots of $R_{\tilde {\dot 1}}$ and 
$R'_{\tilde {\dot 1}}$. Whereas this does not necessarily mean that all roots coincide, we will assume $R_{\tilde 1}=R'_{\tilde 1}$
and $R_{\tilde {\dot 1}}=R'_{\tilde {\dot 1}}$. In this case we get a nice cancellation in the above equation, which further supports this requirement.
Then we get a simple relation
\beq\label{eq:AtildeD}
({\mathcal{A}} )^{\gamma} \, \dot{\mathcal{A}}
= \left(\frac{ R_{2,(+)} }{R_{2,(-)} } \right) \left( \bar{f}_2^{--} f^{++}_2 \bar{ f}_{\dot 2}^{--} {f}_{\dot 2}^{++}\right)\;.
\eeq
It is striking to compare this with the consequence  of the second relation in (\ref{eq:P1tildeABA3}), which yields 
\beq\label{eq:DtildeA}
{\mathcal{A}} \, (\dot{\mathcal{A}})^{\gamma} = \left(\frac{ R_{\dot 2, (+)} }{R_{\dot 2, (-)} } \right) \, \left( \bar{f}_2^{--} f^{++}_2 \bar{ f}_{\dot 2}^{--} {f}_{\dot 2}^{++}\right).
\eeq
Now we continue this relation along the reverse path  $\bar\gamma$: the result on the l.h.s. is $(\mathcal{A})^{\bar\gamma}\dot{\mathcal{A}}$, and the analytic continuation of the r.h.s. is simple to compute, since the $f^{++}$, $\bar f^{ --}$ functions have no cut on the real axis, so are left unchanged. By comparing the result with (\ref{eq:AtildeD}), we get the following ``double-discontinuity" relations
\beq\la{doubledisc}
\frac{\mathcal{A} ^{\gamma}}{ \mathcal{A} ^{\bar\gamma}} = \frac{ R_{2,(+)} }{R_{2,(-)} }  \frac{B_{\dot 2, (-)} }{ B_{\dot 2, (+)} }\,,\qquad\qquad
\frac{\mathcal{\dot A} ^{\gamma}}{ \mathcal{\dot A} ^{\bar\gamma}} = \frac{ R_{\dot 2,(+)} }{R_{\dot 2,(-)} }  \frac{B_{2, (-)} }{ B_{2, (+)} }\;
,
\eeq
where the r.h.s. clearly cannot vanish (except for the vacuum) since the $R_{\alpha}$ and $B_{\alpha}$ functions have zeros on different sheets. 
We will now solve \eq{eq:AtildeD}
and \eq{eq:DtildeA}.

\subsubsection{Relation to the dressing phases}\label{sec:dressingrel}
In order to find the solution, without lack of generality we introduce 
the following ansatz in terms of 
$\rho$ and $\dot \rho$
\beqa
\mathcal{A}= \sqrt{\frac{B_{2,(+)}}{B_{2,(-)}}} \; \sigma^{1, \text{BES}}_2 \,\sigma^{1,\text{BES}}_{\dot 2} \,
\rho\,,\qquad\qquad \dot{\mathcal{A}} =\sqrt{\frac{B_{\dot2, (+)}}{B_{\dot 2, (-)}}}  \; \sigma^{1,\text{BES}}_2 \,\sigma^{1,\text{BES}}_{\dot 2}  \,
\dot\rho
\;,
\eeqa
where, using notation from \cite{Gromov:2014caa},  $\sigma^{1,{\rm BES}}_\alpha$
 denote natural building blocks of the Beisert-Eden-Staudacher dressing factor. They
 satisfy
\beq
( \sigma^{1,{\rm BES}}_{\alpha} )^{\gamma} \;
\sigma^{1,{\rm BES}}_{\alpha}
=
( \sigma^{1,{\rm BES}}_{\alpha} )^{\bar\gamma} \;
\sigma^{1,{\rm BES}}_{\alpha}
\propto f^{ ++}_{\alpha} \bar{f}^{ --}_{\alpha}
\;\;,\;\;
\alpha=2,\dot2\; ,
\eeq
and are related to the 
 product of the BES dressing factors via
\beq
 \sigma_{\rm BES}(u)=\frac{\sigma^{1,\text{BES}}(u+\tfrac{i}{2}) }{\sigma^{1,\text{BES}}(u-\tfrac{i}{2}) }\; ,
\eeq
with the notation explained in appendix  \ref{app:ABA}. With this redefinition,  \eq{eq:AtildeD}, (\ref{eq:DtildeA}) 
 become 
\beq\label{eq:lastRH}
( \rho )^{\gamma} \dot\rho \propto \sqrt{\frac{R_{2,(+)}}{R_{2,(-)}}\; \frac{B_{\dot 2,(-)}}{B_{\dot 2, (+)}} }\,,\qquad\qquad (\dot\rho )^{\gamma} \,\rho \propto \sqrt{\frac{R_{\dot 2,(+)}}{R_{\dot 2,(-)}}\; \frac{B_{ 2,(-)}}{B_{ 2, (+)}} }\;.
\eeq
In appendix \ref{app:crossing}, we define the functions $\sigma^{1,{\rm extra}}$ and
 $\tilde\sigma^{1,{\rm extra}}$
which are related to the two independent dressing phases appearing in ABA equations of section \ref{sec:ABA} in the following way:
\beq
\sigma(u) = \sigma_{\rm BES}(u) \frac{\sigma^{1,{\rm extra}}(u+\tfrac i2)}{\sigma^{1,{\rm extra}}(u-\tfrac i2)}\,,\qquad\qquad \tilde{\sigma}(u) = \sigma_{\rm BES}(u) \frac{\tilde{\sigma}^{1,{\rm extra}}(u+\tfrac i2)}{\tilde{\sigma}^{1,{\rm extra}}(u-\tfrac i2)}.
\eeq
In the same appendix, we also show that these extra pieces satisfy the following identities
\beqa\label{eq:crossstrange}
( {\sigma}^{1,{\rm extra}}_{\alpha} )^{\gamma} \; {\tilde{\sigma}_{\alpha}}^{1,{\rm extra}} = \sqrt{\frac{R_{\alpha,(+)}}{R_{\alpha,(-)}}}, \;\;\; ( {\tilde{\sigma}}^{1,{\rm extra}}_{\alpha} )^{\gamma} \; {{\sigma}_{\alpha}}^{1,{\rm extra}}= \sqrt{\frac{B_{\alpha,(-)}}{B_{\alpha,(+)}}}\;\;,\;\;\alpha\in \left\{2,\dot{2}\right\} \;,
\eeqa
which we both verify directly and also independently deduce from crossing via functional arguments. 

Using those building blocks, we can write
\beq
\rho = {\sigma}_2^{1,{\rm extra}}{\tilde \sigma}_{\dot 2}^{1,{\rm extra}}\rho_0\;\;,\;\;
\dot \rho = {\sigma}_{\dot 2}^{1,{\rm extra}}{\tilde \sigma}_{2}^{1,{\rm extra}}\dot \rho_0\;,
\eeq
where $\rho_0$ and $\dot\rho_0$ should be functions with square-root branch cut on the real axis satisfying
\beq
\rho^\gamma_0 = 1/\dot\rho_0\;\;,\;\;
\dot \rho^\gamma_0 = 1/\rho_0\;.
\eeq
This equation tells us that $\rho^\gamma_0$
is a function with a single cut and neither zeroes nor poles, and likewise $\dot\rho_0$ and $\rho_0$. In other words it can only be a power of $x$, which can be included into a re-definition of $L$. So without 
reducing the generality we can set $\dot\rho_0=\rho_0=1$. This completes the derivation of the asymptotic limit of our QSC.

\subsection{Summary of results for the asymptotic limit}
Let us summarise what we found for the expressions of $\bP$ and $\bQ$ functions.
In the first wing we have:
\begin{align}
\bP_1 &\propto x^{-L/2} \, R_{\tilde{1} } B_{\tilde{\dot  1}} \,
\sqrt{B_{2, (+)  } B_{2, (-)  } }\;\sigma^{1}_2 \, \tilde{\sigma}^{1}_{\dot 2}, & \bP^2 &\propto x^{-L/2} \, R_{\tilde 3} B_{\tilde{\dot 3}} \sqrt{B_{2, (+)  } B_{2, (-)  } }\;\sigma^{1}_2 \, \tilde{\sigma}^{1}_{\dot 2},\label{eq:Qfirstwingfinal0}\\
\bQ_1  &\propto x^{L/2 } R_{1} B_{\dot 1}
 \,\sqrt{\frac{B_{2,(-)}}{B_{2,(+)}}}\; \frac{f_2\,f_{\dot 2} }{ B_{\dot 2, (+) } \, \sigma^{1}_2 \, \tilde{\sigma}^{1}_{\dot 2}}, 
 &
\bQ^2 & \propto  x^{{L}/2}  R_{3} B_{\dot 3} \,\sqrt{\frac{B_{2,(-)}}{B_{2,(+)}}} \; \frac{f_2\,f_{\dot 2} }{ B_{\dot 2, (+) } \, \sigma^{1}_2 \, \tilde{\sigma}^{1}_{\dot 2}},\label{eq:Qfirstwingfinal}
\end{align}
and in the second wing:
\begin{align}
\bP_{\dot 1} &\propto x^{-{L}/2} \, R_{\dot 3} B_{3} \, \sqrt{B_{\dot 2,(+)} B_{\dot 2,(-)}}\; \;\sigma^{1}_{\dot 2} \, \tilde{\sigma}^{1}_{ 2} ,
& \bP^{\dot 2} & \propto  x^{-{L}/2} \, R_{\dot 1} B_{1} \,\sqrt{B_{\dot 2,(+)} B_{\dot 2,(-)}}\; \;\sigma^{1}_{\dot 2} \, \tilde{\sigma}^{1}_{ 2}, \label{eq:Qsndwingfinal0}
\\
\bQ_{\dot 1}  &\propto {x^{{L}/2} }\,  R_{\tilde{\dot  3} } B_{\tilde 3 }  \,\sqrt{\frac{  B_{\dot 2, (-)} }{  B_{\dot 2, (+)}  }} \, \frac{f_{2} f_{\dot 2} }{B_{2, (+) } \, \sigma^{1}_{\dot 2} \, \tilde{\sigma}^{1}_{ 2} } , 
&\bQ^{\dot 2} &\propto {x^{{L}/2} }\, R_{\tilde{\dot 1}} B_{\tilde 1} \,\sqrt{\frac{  B_{\dot 2, (-)} }{  B_{\dot 2, (+)}  }} \, \frac{f_{2} f_{\dot 2} }{B_{2, (+) } \, \sigma^{1}_{\dot 2} \, \tilde{\sigma}^{1}_{2} } .\label{eq:Qsndwingfinal}
\end{align}
All relevant notations are collected in Appendix \ref{app:ABA}. 
Having an asymptotic solution for all relevant $\bP$ and $\bQ$ functions we can plug them into the exact Bethe ansatz equations \eq{eq:ABA1L}-\eq{eq:ABA3Lb} and compare the result with the ABA \eq{ABA1}.

\subsection{Match with the Asymptotic Bethe Ansatz}

We have finally arrived at a full specification of the Q-functions $\bP_1$, $\bP^2$, $\bQ_1$, $\bQ^2$, $Q_{1|1}$, and their dotted cousins, in the ABA limit. To obtain the Asymptotic Bethe Ansatz, we can just plug their values in the exact Bethe equations \eq{eq:ABA1L}-\eq{eq:ABA3Lb} following from the Q-system. 

We will have the following correspondence between the zeros appearing on the first sheet of the Q-functions, and the Bethe roots appearing in the Asymptotic Bethe Ansatz: for the first wing
\beq
\begin{array}{cccc}
\text{Roots:} & u_{1,k} & u_{2,k} & u_{3,k} \\
\text{Q-function:} & \bQ_{1} & Q_{1|1} & \bQ^2 
\end{array} ,\;\;\;\;\; \begin{array}{ccc}
\text{Dual roots:} & {u}_{\tilde{1},k} &{u}_{\tilde{3},k} \\
\text{Q-function:} & \bP_{1} &  \bP^2 
\end{array},
\eeq
and for the second wing:
\beq
\begin{array}{cccc}
\text{Root:} & u_{\dot 1,k} & u_{\dot2, k}& u_{\dot 3, k} \\
\text{Q-function:} & {\bP}_{ \dot 1} & {Q}_{\dot 1|\dot 1} & \bP^{\dot 2}
\end{array}, \;\;\;\;\; \begin{array}{ccc}
\text{Dual roots:} & {u}_{\tilde{\dot 1},k} &{u}_{\tilde{\dot 3},k} \\
\text{Q-function:} & \bQ_{\dot 1} &  \bQ^{\dot 2 } 
\end{array}.
\eeq

In particular, the exact Bethe equations (\ref{eq:ABA1L})-(\ref{eq:ABA3L}) for the first wing 
 reduce precisely the ABA equations (\ref{eq:Bethe13})-(\ref{eq:Bethe33}) using the Q-functions  (\ref{eq:Qfirstwingfinal0})-(\ref{eq:Qfirstwingfinal}). 
 Similarly, using the exact Bethe equations of the form (\ref{eq:ABA1Lb})-(\ref{eq:ABA3Lb}), but for the dotted Q-functions, we reproduce the ABA equations (\ref{eq:Bethe21})-(\ref{eq:Bethe23}) using the asymptotic values (\ref{eq:Qsndwingfinal0})-(\ref{eq:Qsndwingfinal}).  
 
As an example to demonstrate the procedure, we display the case of the middle-node equation for the first wing. At the roots of $Q_{1|1}$ i.e. at $u=u_{2,i}$ we have
\begin{eqnarray}
&&-1 = \frac{Q_{1|1}^{++} Q_{\emptyset|1}^- Q_{12|1}^-}{Q_{1|1}^{--} Q_{\emptyset|1}^+ Q_{12|1}^+}
= \frac{Q_{1|1}^{++} \bQ_{1}^- \bQ^{2-}}{Q_{1|1}^{--} \bQ_{1}^+ \bQ^{2+}}
\nonumber\\
&&\qquad \qquad  = \frac{(x^-)^L (\sigma_2^{1+}\tilde{\sigma}_{\dot{2}}^{1+})^2 {\mathbb Q}_2^{++}  R_1^- B_{\dot{1}}^-  R_3^- B_{\dot{3}}^- 
f_2^{[+3]}f_{\dot{2}}^{[+3]}f_2^-f_{\dot{2}}^- 
B_{2(-)}^-   B_{2(+)}^+ [B_{\dot{2}(+)}^+]^2}
{
 (x^+)^L (\sigma_2^{1-}\tilde{\sigma}_{\dot{2}}^{1-})^2
 {\mathbb Q}_2^{--}
 R_1^+ B_{\dot{1}}^+R_3^+ B_{\dot{3}}^+ f_2^+\;\;\; f_{\dot{2}}^+\;\;\;f_2^+f_{\dot{2}}^+B_{2(-)}^+B_{2(+)}^- [B_{\dot{2}(+)}^-]^2    },\nonumber
\end{eqnarray}
where we have cancelled some terms repeated in the numerator and denominator. 
Next we have to use the defining property of the function $f_\alpha$:
$\frac{f_\alpha^{++}}{f_\alpha} = \frac{B_{\alpha,(-)}}{B_{\alpha,(+)}}$
in appropriate shifted version, to re-create various $B$ functions, some of which then cancel out and some remain. At the end of this massive simplification what is left is exactly the middle-node ABA equation for the first wing, where one needs to recall how the dressing phases are reconstructed from $\sigma^1$ and $\tilde{\sigma}^1$ via $\sigma_\alpha =\frac{\sigma^{1+}}{\sigma^{1-}}$ and $\tilde\sigma_\alpha =\frac{\tilde\sigma^{1+}_{\alpha}}{\tilde\sigma^{1-}_{\alpha}} $: 
\begin{eqnarray}
-1 &=&  \left(\frac{x^{-} }{x^{+}}\right)^{L} \times \frac{\mathbb{Q}^{++}_{2}}{\mathbb{Q}^{--}_{2} }\times \left(\sigma_{2} \right)^2 \times \frac{R_{1}^- R_{3}^- }{ R_{1}^+ R_{3}^+ }\nonumber\\
&&\times \left. \frac{B_{\dot 2,(-) }^+ \, B_{\dot 2, (+) }^+}{B_{\dot 2 , (-) }^- \, B_{\dot 2, (+)  }^-}\times \left( {\tilde\sigma}_{\dot 2} \right)^2\times \frac{B_{\dot 1}^- B_{\dot 3}^- }{ B_{\dot 1}^+ B_{\dot 3}^+ } \right|_{u = u_{2,i}},\,\; i=1,\dots, K_2.
\end{eqnarray}

Finally, since the ABA equations \eq{ABA1}
in the classical regime,  $h\rightarrow \infty$, $L\sim K_\alpha\rightarrow \infty$ with fixed $\mathcal{L}\equiv L/h$
reproduce 
 the classical limit (\ref{eq:classical1})-(\ref{eq:classical4})
 via condensation of roots into cuts in the standard way~\cite{Babichenko:2014yaa}, it follows that we also reproduce the classical limit from the QSC, similarly to \cite{Gromov:2014caa}.  Thus we see that our QSC successfully reproduces all the data from section~\ref{sec:data}.

\section{Discussion and outlook}
The QSCs for $AdS_5$
and $AdS_4$ have a lot in common with one another -- both are based on QQ-relations dictated by the global symmetries and have similar additional analyticity constraints. We use these general features to propose a QSC for
string theory on $AdS_3 \times S^3 \times T^4$ with RR charge and its $\CFT_2$ dual. However, in contrast to the higher-dimensional QSCs, the assumption of square-root singularity near the branch points needs to be dropped. While we reproduced successfully the ABA equations for massive modes, we should still emphasise that, unlike in the previous cases~\cite{Bombardelli:2017vhk,Gromov:2014caa}, we do not have the luxury of TBA equations which can be used as a starting point to derive the QSC equations. Instead, we use a bottom-up approach where  we guess the QSC based on the symmetries and analogy with previous cases, and then verify it in some limits.

On the important  point of the order of the cuts, we notice that if we assume them to be of the usual square-root type, unlike in previous cases, we get a further nontrivial algebraic constraint \eq{doubletrouble} on the Q-functions in addition to the QQ-relations, resulting in a too restrictive set of equations. So to some extent the absence of square-root behaviour is dictated by  symmetries.

To be fully confident in the self-consistency and completeness of equations proposed here,  we need to perform further tests beyond the matching with the ABA presented here. For example, constructing the perturbative weak coupling solution at several loop orders would be useful, which can be done with the methods of \cite{Gromov:2015vua,Marboe:2014gma}. The QSC should also reproduce the protected spectrum of the theory~\cite{deBoer:1998kjm}, accounting for all-order wrapping corrections not considered in the ABA analysis~\cite{Baggio:2017kza,Majumder:2021zkr}. Further, it would be interesting to consider  near-BPS limits where one can expect a non-trivial analytic solution at finite coupling ~\cite{Gromov:2014bva,Gromov:2014eha}. Finally, one should try to solve the system numerically with high precision like in \cite{Gromov:2015wca}.
Another potential way to test our equations would 
be to re-derive the TBA equations for the massless modes \cite{Bombardelli:2018jkj,Fontanella:2019ury}.

An important question to address is whether the massless modes of the theory are already contained in our QSC proposal or whether the construction needs to be generalised in some way. For example, one might wonder whether it is possible to take a tensor product of our Q-system with an additional Q-system, perhaps based on $\alg{su}(2)_\circ$ under which the massless bosons are known to transform~\cite{Borsato:2013qpa}. Unfortunately, such a direct product is not compatible with the fact that massless fermions transform non-trivially under the $\alg{psu}(1,1|2)^2$ symmetry. On the other hand, the structure of the Q-system is rather rigid so it is harder to see how to augment it by an additional Q-system in a more non-trivial way.

Alternatively, it may be that incorporating the massless modes requires relaxing slightly some of the analyticity and pole-structure properties we require of the QSC. The starting point for this approach would be an attempt to  derive the ABA equations including massless modes in the large-volume limit, generalising the construction we presented above. This would require new arguments, since the notion of the asymptotic regime is delicate in the presence of massless excitations, where the standard exponential large-$L$ suppression is no longer there and wrapping corrections often are of the same order as the ABA contributions. 
However, we point out that the QSC structure is typically very rigid and does not allow for much more freedom. In particular, given the underlying $\alg{psu}(1,1|2)$ algebraic structure of our system, we believe that the only place where the QQ relations could be changed would be a modification of the condition $Q_{12|12}=Q_{\dot 1\dot 2|\dot 1\dot 2}=1$. We might also have to modify the gluing condition, although this seems less natural. 
 However, we cannot at this stage exclude the possibility that neither of these options will be necessary and the QSC presented here is already complete. 
 To provide evidence for this claim, it would be very interesting to solve the QSC equations at finite coupling, as this would allow to perform tests in various limits, for example comparing with massless solutions at weak coupling or in the semi-classical regime~\cite{Varga:2020umx}.


If these additional tests can be satisfactorily performed, one can hope that $AdS_3$ would become an ideal background for application of SoV program for correlators~\cite{Cavaglia:2018lxi}. Further, combining the $AdS_3$ QSC  spectral methods with Conformal Bootstrap~\cite{Cavaglia:2021bnz} techniques could provide a simpler testing ground for these ideas compared to  the  ${\cal N}=4$ SYM case.

Following these tests of our conjecture, it would be interesting to extend the $AdS_3$ QSC construction to $AdS_3 \times S^3 \times T^4$ backgrounds supported by combinations of RR and NSNS charges. The ABA for these theories is also known~\cite{Lloyd:2014bsa} and solutions to the crossing equations have recently been found~\cite{OSS-Stef-to-appear}, which should provide a further testing ground for the QSC analysis. String theory on $AdS_3 \times S^3 \times S^3\times S^1$ is also expected to be integrable~\cite{Borsato:2015mma}. Finding the QSC for this model would be particularly interesting since the global symmetry algebra is $d(2,1;\alpha)^2$, for which the Q-system should exhibit novel features.

It would be interesting to see whether similar techniques to the ones we have employed here can be extended to the $AdS_2/CFT_1$ integrable system~\cite{Hoare:2014kma}, which also features the presence of massless modes and has an algebraic structure of a similar complexity. The issue of long vs short representations, which is relevant in that case, is likely to represent an additional novelty and a reason for adapting the method even further.  

\section{Acknowledgments}
B.S. and A.T. would like to thank Alessandro Sfondrini and Dima Volin for useful discussions during the ``Integrability  in  Lower  Dimensional  AdS/CFT" workshop.
A.C. and N.G. would like to thank Julius, Vladimir Kazakov and Fedor Levkovich-Maslyuk for inspiring conversations.
B.S. acknowledges funding support
from an STFC Consolidated Grant ‘Theoretical Particle Physics at City, University of London' ST/T000716/1.
A.T. is supported by the EPSRC-SFI grant EP/S020888/1 {\it Solving Spins and Strings}.
The work of A.C. and N.G. is supported by European Research Council (ERC) under the European Union’s Horizon 2020 research and innovation programme (grant agreement No. 865075)  EXACTC.
N.G. is also partially supported by the STFC grant (ST/P000258/1). During the first phase of this work, A.C. was supported by the STFC grant (ST/P000258/1). We thank the Hamilton Mathematical Institute (Dublin) for hosting the workshop ``Integrability in Lower Dimensional AdS/CFT", which was a catalyst for the final stages of this work.

\appendix

\section{Rewriting the ABA equations}\label{app:ABA}
\paragraph{Notations. }
We introduce some useful notations for the ABA equations. First, using the Zhukovsky map, 
\beq
x(u) = \frac{1}{2 h}\left(u + \sqrt{ u - 2 h}\sqrt{ u + 2 h} \right),
\eeq
we reparametrise the roots in terms of $u_{2,k}$, $u_{1,k}$, $u_{3,k}$, such that $ x^{\pm}_k \equiv x(u_{2, k} \pm \frac{i}{2}) $,  $1\leq k\leq K_{{2}}$, $y_{1,k} \equiv x(u_{1, k})$, $1\leq k\leq K_{{1}}$, $y_{3,k} \equiv x(u_{3, k}) $, $1\leq k\leq K_{{3}}$, and similarly for the other wing introducing  $u_{\dot 2,k}$, $u_{\dot 1, k}$, $u_{\dot 3, k}$. We also accordingly rename 
 $
K_{\bar{\alpha}} \equiv K_{\dot \alpha}$,  $\alpha = 1,2,3$,
 as compared to the notations of section 
 \ref{sec:ABA}. 

It is convenient to introduce the generalised Baxter polynomials
\beqa
B_{\alpha,(\pm)}(u) &\equiv& \prod_{j=1}^{K_{\alpha}} \sqrt{\frac{h}{x_{\alpha,j}^{\mp}} }(\frac{1}{x(u)} - x_{\alpha,j}^{\mp} ) ,\;\;\;\; \alpha \in\left\{2,\dot 2 \right\}  \\
R_{\alpha, (\pm)}(u) &\equiv& \prod_{j=1}^{K_{\alpha}} \sqrt{\frac{h}{x_{\alpha,j}^{\mp}} }(x(u) - x_{\alpha,j}^{\mp} ) ,\;\;\;\; \alpha \in\left\{2,\dot 2 \right\} \\
\mathbb{Q}_{\alpha}(u) &=& \prod_{j=1}^{K_{\alpha}} (u - u_{\alpha, j} ),\;\;\;\; \alpha \in\left\{2,\dot 2 \right\} ,
\\
B_{\alpha}(u) &=& \prod_{j=1}^{K_{\alpha} } ({\frac{1}{x(u)}} - y_{\alpha, j} ), \;\; \alpha\in  \left\{ 1,3,\dot 1,\dot 3\right\} \label{eq:defB}\\
R_{\alpha}(u) &=& \prod_{j=1}^{K_{\alpha} } ({x(u)} - y_{\alpha, j} ), \;\; \alpha \in \left\{ 1,3,\dot 1,\dot 3\right\}.\label{eq:defR}
\eeqa
Notice that $B_{\alpha, (\pm)} R_{\alpha, (\pm)} \propto \mathbb{Q}_{\alpha}^{\pm}$, where the shift of a function of $u$ is defined as 
 $
g^{[\pm n]}(u)\equiv g(u + i \frac{n}{2})$, $g^{\pm}\equiv g^{[\pm 1]}$. Through the Zhukovsky map, we also consider the dressing phase a function of rapidities:
\beq
\sigma(u, v) \equiv \text{exp}\left( i\chi(x^+(u), x^+(v))-i\chi(x^+(u), x^-(v))+i\chi(x^-(u), x^-(v))-i\chi(x^-(u), x^+(v)) \right),
\eeq
and we introduce the notation:
\beqa
\sigma_{\alpha}(u) &\equiv& \prod_{i=1}^{K_{\alpha}}\sigma(u , u_{\alpha, i} ), \;\;\; \alpha = 2, \dot{2} ,
\eeqa
and the same conventions are taken for $\tilde{\sigma}$. 
We also use the same notation for the BES dressing phase. 
We also introduce useful building blocks
\beq
\sigma^1(u, v) \equiv \text{exp}\left( i\chi(x(u), x^+(v))-i\chi(x(u), x^-(v)) \right),
\eeq
and similarly for $\tilde{\sigma}^1$, and $\sigma^{1\rm BES}$, and 
denote again the products over roots as
\beq 
\sigma^1_{\alpha}(u)\equiv  \prod_{i=1}^{K_{\alpha}} \sigma^1(u, u_{\alpha, i} ),
\eeq
with the analogous definitions made for $\tilde{\sigma}^1_{\alpha}(u)$ and $\sigma^{1,\text{BES}}(u)$. We then have the relation
\beq
\sigma_{\alpha}(u) = \frac{\sigma^1_{\alpha}(u+\frac{i}{2})}{\sigma^1_{\alpha}(u-\frac{i}{2})},
\eeq
and its generalisations. It will also be useful for some of our discussions to define $\sigma^{1,\text{extra}}_{\alpha}(u)$,  $\tilde{\sigma}^{1,\text{extra}}_{\alpha}(u)$ through
\beq
\sigma^1_{\alpha}(u) \equiv  \sigma^{1,\text{BES}}_{\alpha}(u) \; \sigma^{1,\text{extra}}_{\alpha}(u), \;\;\; \tilde{\sigma}^1_{\alpha}(u) \equiv  \sigma^{1,\text{BES}}_{\alpha}(u) \; \tilde{\sigma}^{1,\text{extra}}_{\alpha}(u).
\eeq
Finally, for the reader's convenience we collect the defining relations for the functions $f_{\alpha}$, $\bar{f}_{\alpha}$ appearing in the large-volume solution of the QSC:
\beq
\frac{f_{\alpha}^{++}}{f_{\alpha} } = \frac{B_{\alpha,(-)}}{B_{\alpha,(+)}} , \;\;\; \frac{\bar{f}_{\alpha}}{\bar{f}_{\alpha}^{--} }= \frac{B_{\alpha,(-)}}{B_{\alpha,(+)}},
\eeq
where $f_{\alpha}$ is assumed analytic in the upper half plane and $\bar{f}_{\alpha}$ in the lower half plane, and both are free of poles everywhere. These functions are given explicitly (up to an arbitrary multiplicative constant) by DHM-type integral representations similar to (\ref{eq:intf}).

\paragraph{Compact rewriting of the ABA equations. }

With the notations above, the ABA equations can be rewritten as
\beqa
1 &=& \left. \frac{ \mathbb{Q}^{-}_{2} \, B_{2,(+)} \, B_{\dot 2,(+) } }{ \mathbb{Q}^{+}_{2} \, B_{2,(-) } \, B_{\dot 2,(-) } } \right|_{u = u_{1,  i}}, \,\,\, i=1,\dots, K_{1} , \label{eq:Bethe13} \\
-1 &=&  \left(\frac{x^{[-]} }{x^{[+]}}\right)^{L} \times \frac{\mathbb{Q}^{++}_{2}}{\mathbb{Q}^{--}_{2} }\times \left(\sigma_{2} \right)^2 \times \frac{R_{1}^- R_{3}^- }{ R_{1}^+ R_{3}^+ } \nonumber \\
&\times& \left. \frac{B_{\dot 2,(-) }^+ \, B_{\dot 2, (+) }^+}{B_{\dot 2 , (-) }^- \, B_{\dot 2, (+)  }^-}\times \left( {\tilde\sigma}_{\dot 2} \right)^2\times \frac{B_{\dot 1}^- B_{\dot 3}^- }{ B_{\dot 1}^+ B_{\dot 3}^+ } \right|_{u = u_{2,i}},\,\; i=1,\dots, K_2 \label{eq:ABA2expl}\\
1 &=& \left. \frac{ \mathbb{Q}^{-}_{2} \, B_{2,(+)} \, B_{\dot 2,(+) } }{ \mathbb{Q}^{+}_{2} \, B_{2,(-) } \, B_{\dot 2,(-) } } \right|_{u = u_{3,  i}}, \,\,\, i=1,\dots, K_{3} \label{eq:Bethe33}
\eeqa
for the first wing, and
\beqa
1 &=& \left. \frac{ \mathbb{Q}^{-}_{\dot 2} \, B_{\dot 2,(+)} \, B_{  2,(+) } }{ \mathbb{Q}^{+}_{\dot 2} \, B_{\dot 2,(-) } \, B_{  2,(-) } } \right|_{u = u_{\dot 1,  i}}, \,\,\, i=1,\dots, K_{\dot 1} , \label{eq:Bethe21} \\
-1 &=& \left(\frac{{x}^{[-]} }{{x}^{[+]}}\right)^{-L} \times \frac{\mathbb{Q}^{++}_{\dot 2}}{\mathbb{Q}^{--}_{\dot 2} }\times \left(\frac{ B_{\dot 2,(-)}^- }{ B_{\dot 2,(+) }^+} \right)^2 \, \left(\sigma_{\dot 2} \right)^{-2}  \times \frac{R_{\dot 1}^- R_{\dot 3}^- }{ R_{\dot 1}^+ R_{\dot 3}^+ } \nonumber \\
&\times& \left. \frac{B_{ 2, (-) }^- \, B_{2,(-) }^+}{B_{  2,(+) }^- \, B_{ 2,(+) }^+}\times \left(\tilde{\sigma}_{ 2} \right)^{-2} \times \frac{B_{1}^- B_{3}^- }{ B_{1}^+ B_{3}^+ } \right|_{u = u_{\dot 2, i}},\,\; i=1,\dots, K_{\dot 2} \label{eq:ABA2bexpl}\\
1 &=& \left. \frac{ \mathbb{Q}^{-}_{\dot 2} \, B_{\dot 2,(+)} \, B_{  2,(+) } }{ \mathbb{Q}^{+}_{\dot 2} \, B_{\dot 2,(-) } \, B_{  2,(-) } } \right|_{u = u_{\dot 3,  i}}, \,\,\, i=1,\dots, K_{\dot 3} \label{eq:Bethe23}
\eeqa
for the second wing. 

\section{\label{app:crossing} Functional equations for the building blocks of dressing factors }\label{eq:crossing}

In this section, we decompose the two types of dressing factors appearing in the ABA as
\begin{equation}
\sigma(u,v)= \sigma_{\rm BES}(u,v)\,\sigma_{\rm extra}(u,v)\,,\qquad
\tilde{\sigma}(u,v) = \sigma_{\rm BES}\,\tilde{\sigma}_{\rm extra}(u,v)\,,
\end{equation}
and similarly
\beq
\sigma^1(u,v)= \sigma^{1,\rm BES}(u,v)\,\sigma^{1,\rm extra}(u,v)\,,\qquad
\tilde{\sigma}^1(u,v) = \sigma^{1,\rm BES}\,\tilde{\sigma}^{1,\rm extra}(u,v) ,
\eeq
see section \ref{app:ABA} for notation. The goal of this appendix is to establish the functional relations 
\beq\label{eq:firsttarget}
\begin{aligned}
{\sigma^{1,\rm extra}}( u^{\gamma}, v) \;{\tilde{\sigma}^{1,\rm extra}}( u, v)  =\sqrt{\frac{R_{(+)}(u,v)}{R_{(-)}(u,v)}}\,,
\\
{\tilde{\sigma}^{1,\rm extra}}( u^{\gamma}, v)  \;{{\sigma}^{1,\rm extra}}( u, v)  =\sqrt{\frac{B_{(-)}(u,v)}{B_{(+)}(u,v)} } \,,
\end{aligned}
\eeq
where in this appendix we denote 
\beq
R_{(\pm)}(u, v) = \frac{x(u) - x^{\mp}(v) }{\sqrt{x^{\mp}(v)}} , \;\;\; B_{(\pm)}(u, v) = R_{(\pm)}(u^{\gamma}, v) = \frac{\frac{1}{x(u)} - x^{\mp}(v) }{\sqrt{x^{\mp}(v)}} .
\eeq
These relations are important for deriving the ABA from the QSC, as they imply the crucial equation  (\ref{eq:crossstrange}).  In presenting their proof here, we will also deduce
\beq
\begin{aligned}\label{eq:secondtarget}
{\sigma^{1,\rm extra}}( u^{\gamma^{-1}}, v) \;{\tilde{\sigma}^{1,\rm extra}}( u, v)  =\sqrt{\frac{R_{(-)}(u,v)}{R_{(+)}(u,v)}}\,,
\\
{\tilde{\sigma}^{1,\rm extra}}( u^{\gamma^{-1}}, v)  \;{{\sigma}^{1,\rm extra}}( u, v)  =\sqrt{\frac{B_{(+)}(u,v)}{B_{(-)}(u,v)} } \,.
\end{aligned}
\eeq
\subsection{Direct derivation}
We start by verifying these relations directly, based on the expressions for the dressing phases of  \cite{Borsato:2013hoa}. 
From the results of this paper we deduce
\begin{equation}
\sigma^{1,{\rm extra}}(x,x_2^\pm) = \exp i \Lambda^{LL}(x,x_2^\pm), \qquad \tilde{\sigma}^{1,{\rm extra},RL}(x,x_2^\pm) = \exp i \Lambda^{RL}(x,x_2^\pm),
\end{equation} 
where we have defined\footnote{We use the notation $\chi^-$ of \cite{Borsato:2013hoa}, where the minus does not denote a shift in the spectral parameter but is just a label.}
\begin{eqnarray}
&&\Lambda^{LL}(x,x_2^\pm) = - \frac{1}{2} \chi^{\rm HL}(x,x_2^+) + \frac{1}{2} \chi^{\rm HL}(x,x_2^-)  + \frac{1}{2} \chi^-(x,x_2^+) - \frac{1}{2} \chi^-(x,x_2^-), \nonumber \\
&&\Lambda^{RL}(x,x_2^\pm) = - \frac{1}{2} \chi^{\rm HL}(x,x_2^+) + \frac{1}{2} \chi^{\rm HL}(x,x_2^-)  - \frac{1}{2} \chi^-(x,x_2^+) + \frac{1}{2} \chi^-(x,x_2^-),
\end{eqnarray}
and we have the integral representations
\begin{eqnarray}
&&\chi^{\rm HL}(x,y) = \frac{\pi}{2}\oint \frac{dw}{2\pi i}\oint \frac{dw'}{2\pi i} \frac{\mbox{sign}(w' +1/w' - w - 1/w)}{(x-w)(y-w')},\nonumber\\
&&\chi^-(x,y) = \Bigg(\int_{C^+} - \int_{C^-}\Bigg) \frac{dw}{8\pi} \frac{1}{x-w} \log \Big[ (y-w)\Big(1-\frac{1}{y w}\Big)\Big] - x \leftrightarrow y,
\end{eqnarray}
where the full circles run counterclockwise and the contours $C^\pm$ denote the upper (resp., lower) half semicircle in the complex $w$-plane, running counterclockwise.

We can write 
\begin{eqnarray}
t(x,x_2^\pm) \equiv \Big(\sigma^{1,{\rm extra}}\Big)^2(x,x_2^\pm) = \exp \Big[2 i \chi_{\rm extra}(x,x_2^+) - 2i \chi_{\rm extra}(x,x_2^-)\Big],
\end{eqnarray}
and the same for the other block denoted with tilde, where 
\begin{eqnarray}
\chi_{\rm extra}=-\frac{1}{2}\Big(\chi^{\rm HL}-\chi^-\Big), \qquad \tilde{\chi}_{\rm extra}=-\frac{1}{2}\Big(\chi^{\rm HL}+\chi^-\Big),
\end{eqnarray}
and $\chi^{\rm HL}$ and $\chi^-$ are given by an explicit integral representation. 

Equations (3.8) and (3.9) of~\cite{Borsato:2013hoa} are consistent with 
\begin{eqnarray}
\chi^{\rm HL} + ({\chi}^{\rm HL} )^{\gamma^{-1}} = \frac{i}{2} \log \ell^{\rm HL}, \qquad \ell^{\rm HL}(x,y) = \frac{x-y}{1-{xy}}, 
\end{eqnarray}
while equations (3.14) and (3.15) in the same paper are consistent with 
 \begin{eqnarray}
\chi^- - ({\chi}^- )^{\gamma^{-1}}= \frac{i}{2} \log \ell^-, \qquad \ell^-(x,y) = (x-y)(1-\frac{1}{xy}). 
\end{eqnarray}
Therefore, we can assemble 
\begin{eqnarray}
({\sigma}^{1,{\rm extra}} )^{\gamma^{-1}}\; \tilde{\sigma}^{1,{\rm extra}} = \exp \frac{1}{4} \log \frac{\ell^{\rm HL}(x,x_2^+)\ell^-(x,x_2^+)}{\ell^{\rm HL}(x,x_2^-)\ell^-(x,x_2^-)} = \Big(\frac{x_2^-}{x_2^+}\Big)^{\frac{1}{4}} \sqrt{\frac{(x-x_2^+)}{(x-x_2^-)}}.
\end{eqnarray}
Recalling the definition of the function $R_{(\pm)}$, we can reproduce the first equation in (\ref{eq:secondtarget}). 
Likewise, we can compute 
\begin{eqnarray}
(\tilde{\sigma}^{1,{\rm extra}} )^{\gamma^{-1}}\; \sigma^{1,{\rm extra}} = 
\exp \frac{1}{4} \log \frac{\ell^{\rm HL}(x,x_2^+)\ell^-(x,x_2^-)}{\ell^{\rm HL}(x,x_2^-)\ell^-(x,x_2^+)} = \Big(\frac{x_2^+}{x_2^-}\Big)^{\frac{1}{4}} \sqrt{\frac{\frac{1}{x} - x_2^-}{\frac{1}{x} - x_2^+}},
\end{eqnarray}
which reproduces the second equation in (\ref{eq:secondtarget}) if we recall the definition of the function $B_{(\pm)}$. 
The other relations in (\ref{eq:constrapp2}) also follow: since the cut is of logarithmic type, we get the reciprocal results on the r.h.s. if we cross it in the other direction.

\subsection{Functional argument }
\begin{figure}[t]
    \centering
    \includegraphics[scale=0.4]{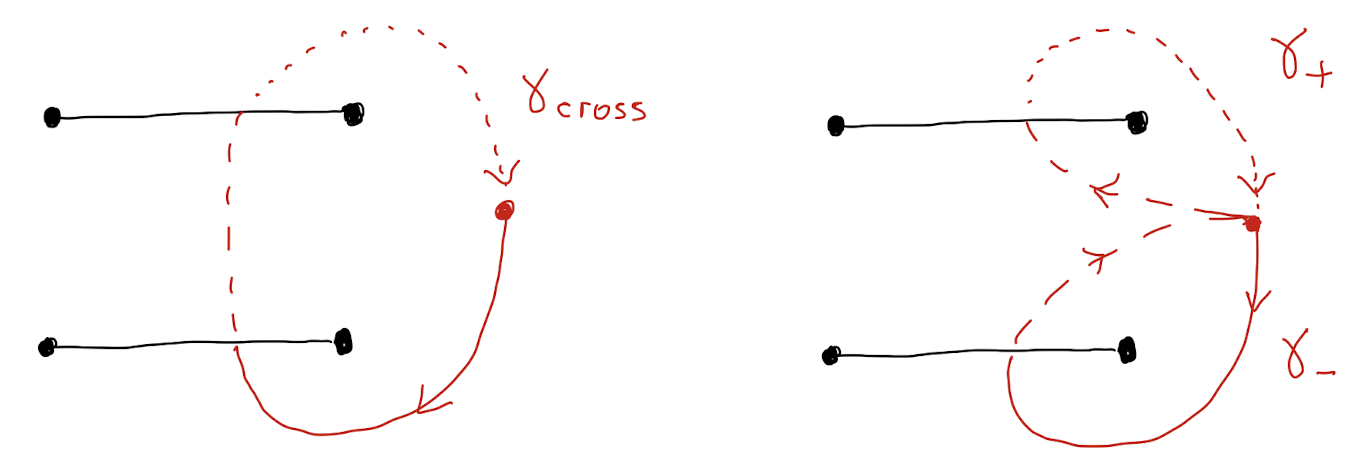}
    \caption{The analytic continuation path used in the crossing relation. It crosses the cuts of the dressing factors at $(-2h, 2h)\pm \frac{i}{2}$, and can be decomposed into $\gamma_{+}$ and $\gamma_-$, which cross only one cut each. }
    \label{fig:gammacross}
\end{figure}

Here we establish the same relations starting from the crossing equation, and assuming certain minimality requirements  on its solution.
The crossing equation can be decomposed into the crossing satisfied by the BES part,
\begin{equation}
\sigma_{\rm BES}(u_1^{\gamma_{\rm cross}},u_2)\sigma_{\rm BES}(u_1,u_2)
=
\frac{x_2^-}{x_2^+}
\frac{x_1^--x_2^+}{x_1^--x_2^-}
\frac{1-\frac{1}{x_1^+x_2^+}}
{1-\frac{1}{x_1^+x_2^-}} ,
\end{equation}
and the crossing relations  for the extra pieces:
\begin{align}\label{eq:extracrossing}
\sigma_{\rm extra}(u_1^{\gamma_{\rm cross}},u_2)^2\,\tilde{\sigma}_{\rm extra}(u_1,u_2)^2
&=
\frac{(x_1^+-x_2^+)(x_1^--x_2^-)}{(x_1^--x_2^+)(x_1^+-x_2^-)},
\\
\sigma_{\rm extra}(u_1,u_2)^2\,\tilde{\sigma}_{\rm extra}({u}_1^{\gamma_{cross}},u_2)^2
&=
\frac{\left(1-\frac{1}{x_1^+x_2^-}\right)\left(1-\frac{1}{x_1^-x_2^+}\right)}
{\left(1-\frac{1}{x_1^+x_2^+}\right)\left(1-\frac{1}{x_1^-x_2^-}\right)} .\label{eq:extracrossing2}
\end{align}
The path $\gamma_{\rm cross}$ is depicted in figure \ref{fig:gammacross},
and it can be decomposed as the concatenation of the path $\gamma_-$, entering the lower cut, followed by $\gamma_+$ which enters the upper cut. 
Notice that in our notations, different from some of the literature, both $\gamma^{\pm}$  cross one cut from below. 

We now follow a similar route to the one described in \cite{Vieira:2010kb}, and disentangle the  path $\gamma_{\rm cross}$ to derive a simpler equation for a natural building block of the solution to the  crossing constraints.  
We will assume that, for the minimal solution, the crossing path is equivalent to the one obtained by concatenating $\gamma_-$ and $\gamma_+$ in opposite order, $\gamma_{\rm cross} \simeq \gamma_+\cdot \gamma_- \simeq \gamma_-\cdot \gamma_+$. Under this assumption, analytically continuing along the inverse path $\gamma_+^{-1}$ the crossing relations (\ref{eq:extracrossing}),(\ref{eq:extracrossing2}), we get:
\beqa
s(u_1^{\gamma_-},u_2)\hat{s}(u_1^{\gamma_+^{-1}},u_2) &=& \frac{R_{(-)}^+ B_{(+)}^- }{R_{(+)}^+ B_{(-)}^-} \equiv A , \label{eq:eqA}\\
\hat{s}(u_1^{\gamma_-},u_2)s(u_1^{\gamma_+^{-1}},u_2) &=& \frac{R_{(-)}^- B_{(+)}^+ }{R_{(+)}^- B_{(-)}^+} \equiv C ,
\eeqa
while continuing the same variable along  $\gamma_-^{-1}$, we get:
\beqa
s(u_1^{\gamma_+},u_2)\hat{s}(u_1^{\gamma_-^{-1}},u_2) &=& \frac{B_{(-)}^+ R_{(+)}^- }{B_{(+)}^+ R_{(-)}^-} \equiv B ,\label{eq:eqB} \\
\hat{s}(u_1^{\gamma_+},u_2)s(u_1^{\gamma_-^{-1}},u_2) &=& \frac{B_{(-)}^- R_{(+)}^+ }{B_{(+)}^- R_{(-)}^+} \equiv D ,
\eeqa
where for simplicity of the next expressions, we denoted $s(u_1,u_2)\equiv \sigma_{\rm extra}(u_1,u_2)^2$,  $\tilde{s}(u_1,u_2)\equiv \tilde{\sigma}_{\rm extra}(u_1,u_2)^2$. 

From now on, we omit the second variable, since it is simply a spectator in all these functional relations, and use the notation $g^{[n]}$, described in the main text, to shift the first variable of various functions. 
We proceed by making the ansatz 
\beq
s\equiv\frac{t^+}{t^-}, \;\;\; \tilde{s}\equiv\frac{\tilde{t}^+}{\tilde{t}^-},
\eeq
where $t$, $\tilde{t}$ are assumed to be functions with a single cut $(-2h, 2h)$. The relations between these blocks and the ones introduced above is simply $t \propto (\sigma^{1,\rm extra} )^2$, $\tilde{t} \propto (\tilde{\sigma}^{1,\rm extra} )^2$. 

Now we notice that, 
\beq
\left(\frac{t^+}{t^-}\right)^{\gamma_-} = \frac{(t^{\gamma^{-1} } )^+}{t^-} \,,\qquad\qquad \left(\frac{t^+}{t^-}\right)^{\gamma_+} = \frac{t^+}{(t^{\gamma^{-1} } )^-},
\eeq
and there are similar relations if we do analytic continuations along the inverse paths $\gamma_{\pm}^{-1}$, which are simply obtained by replacing $\gamma^{-1}\rightarrow \gamma$ on the r.h.s. 
Taking the product of (\ref{eq:eqA}),(\ref{eq:eqB}), we arrive at 
\beq
\left(({t})^{\gamma^{-1}} \;\tilde{t}\; t\; (\tilde{t})^{\gamma }   \right)^{\hat{D}-\hat{D}^{-1}} = A B = \left( \frac{R_{(-)} B_{(-)}}{R_{(+)} B_{(+)}}\right)^{\hat{D}-\hat{D}^{-1}} ,
\eeq
where $\hat{D}\equiv \frac{i}{2} \partial_u$, so that in this notation $g^{n \hat{D}} \equiv g^{[n]}$.  Since we look for the minimal solution to crossing, we take the simplest solution to the previous functional relation:
\beq\label{eq:constrapp1}
({t})^{\gamma^{-1}} \;\tilde{t}\; t\; (\tilde{t})^{\gamma }  =   \frac{R_{(-)} B_{(-)}}{R_{(+)} B_{(+)}}. 
\eeq
Similarly, considering the ratio of  the same two equations, and assuming the minimal solution, we obtain 
\beq\label{eq:constrapp2}
\left( {({t})^{\gamma^{-1}} \;\tilde{t}}\right)/\left({ t\; (\tilde{t})^{\gamma } }\right) =   \frac{R_{(-)} B_{(+)}}{R_{(+)} B_{(-)}},
\eeq
and finally from (\ref{eq:constrapp1}),(\ref{eq:constrapp2})  we read:
\beq
({t})^{\gamma^{-1}} \;\tilde{t} =\frac{R_{(-)}}{R_{(+)}} \,,\qquad\qquad {t} \;(\tilde{t})^{\gamma} =\frac{B_{(-)}}{B_{(+)}} . 
\eeq
By the same arguments from the remaining two equations we extract:
\beq
({t})^{\gamma} \;\tilde{t} =\frac{R_{(+)}}{R_{(-)}} \,,\qquad\qquad ({t}) \;(\tilde{t})^{\gamma^{-1}} =\frac{B_{(+)}}{B_{(-)}} . 
\eeq
Taking into account that, in the notations of the main text, $t \equiv (\sigma^{1,{\rm extra}})^2$, $\tilde{t}\equiv (\tilde{\sigma}^{1,{\rm extra}})^2$, we have therefore deduced the relations (\ref{eq:firsttarget}), (\ref{eq:secondtarget}). 

\section{Baxter equations}\label{app:Baxter}
\paragraph{Baxter equations for $\bQ$ and $\bP$ functions.} 
The obvious identities
\beq
\bQ_k^{++}
\epsilon^{ij}
\bQ_i^{--}\bQ_j
-
\bQ_k
\epsilon^{ij}
\bQ_i^{--}\bQ^{++}_j
+
\bQ^{--}_k
\epsilon^{ij}
\bQ_i\bQ^{++}_j=0\;\;,\;\;k=1,2,
\eeq
can be recast as the Baxter equations
\beq
\bQ_k^{++}
D_1^-
-
\bQ_k
D_2
+
\bQ^{--}_k
D_1^+=0\;\;,\;\;k=1,2 ,
\eeq
where the coefficients can also be rewritten in terms of $\bP$ functions using the QQ relations:
\beqa
D_1\equiv \epsilon^{ij}
\bQ_i^-\bQ_j^+&=&
\epsilon_{ab}
\bP^{a-}\bP^{b+}\\
D_2\equiv \epsilon^{ij}
\bQ_i^{--}\bQ_j^{++}&=&
\epsilon_{ab}
\bP^{a--}\bP^{b++}
-
\bP_c \bP^{c--}
\epsilon_{ab}
\bP^{a}\bP^{b++}\nonumber \\
&=&
\epsilon_{ab}
\bP^{a--}\bP^{b++}
-
\bP_c \bP^{c++}
\epsilon_{ab}
\bP^{a}\bP^{b--}
\eeqa
(the last equality follows from $\bP_a \bP^a=0$).
These equations, supplemented by the large-$u$ asymptotics, give a way to compute the $\bQ$ functions starting from the knowledge of the $\bP$ functions. 

There are also equations of the same form, obtained by replacing $\bP\leftrightarrow\bQ$, which may be used to compute the $\bP$ functions starting from the $\bQ$'s. 
\paragraph{Finite difference relations for $Q_{a|i}$. } 
We close this appendix by noticing that also the middle node Q-functions  can be defined as the solutions of a system of finite-difference equations, which are simply obtained from the Q-system. 

One such system of relations is
\beq\label{eq:Qaifinitediff}
Q_{a|i}^+ - Q_{a|i}^- = \bP_a \bP^b Q_{b|i}^+ .
\eeq
These relations can be used to construct $Q_{a|i}$ from the knowledge of the $\bP$ functions. The  solution is specified by  requiring the appropriate asymptotic behaviour, and the region of analyticity.  Solutions analytic in the upper half plane are denoted as $Q^{\downarrow}_{a|i}$.  
The solutions analytic in the lower half plane form an alternative basis of solutions, denoted by $Q_{a|i}^{\uparrow}$. 
The numerical method to compute $Q_{a|i}^{\downarrow}$ and $Q_{a|i}^{\uparrow}$ in terms of the $\bP$ functions is described in \cite{Gromov:2015wca}. 

The two bases of solutions of the same finite-difference equations are related by an $i$-periodic matrix
\beq\label{eq:QupQdown}
{Q}_{a|i}^{\uparrow+} = \Omega_i^{\;\; j} {Q}_{a| j}^{\downarrow+} , \;\;\; {Q}^{a|i\uparrow+} = \Omega^i_{\;\;j} {Q}^{a| j\downarrow+},
\eeq
which imply
\beq\label{eq:defOmegaapp}
\Omega_k^{\;\;l}  = {Q}_{a|k}^{\uparrow+} Q^{a|l \downarrow+},\;\; \;\Omega^k_{\;\;l}= {Q}^{a|k\uparrow+} Q_{a|l\;\;}^{\downarrow+}.
\eeq
 Multiplying the first equation in (\ref{eq:QupQdown}) by $\bP^a$ on the left, we see immediately  that $\Omega$ is the same matrix relating $\bQ^{\uparrow}$ and $\bQ^{\downarrow}$ in (\ref{eq:defOmega}). Similarly, the second equation show that
 \beq
\bQ^{i\uparrow} = \Omega^i_{\;\;j} \bQ^{j\downarrow}.
\eeq 
Since $Q_{a|i}$ has unit determinant and $Q_{a|i} Q^{a|j} = \delta_i^j$, from (\ref{eq:defOmegaapp}) we see immediately that $\Omega$ has unit determinant as well, and
\beq
\Omega^i_{\;\;j} \Omega_k^{\;\;j} = \delta^i_k.
\eeq
Finally, another useful form of (\ref{eq:Qaifinitediff}) is
\beq
Q_{a|i}^+ - Q_{a|i}^- = \bQ_i \bQ^j Q_{a|j}^+ ,
\eeq
which can be used to determine $Q_{a|i}$ from the knowledge of the $\bQ$ functions. 

\bibliographystyle{JHEP}
\bibliography{references}

\end{document}